\documentclass[fleqn,usenatbib]{mnras}

\usepackage{newtxtext,newtxmath}

\usepackage[T1]{fontenc}
	
\DeclareRobustCommand{\VAN}[3]{#2}
\let\VANthebibliography\thebibliography
\def\thebibliography{\DeclareRobustCommand{\VAN}[3]{##3}\VANthebibliography}

\usepackage{graphicx}	
\usepackage{amsmath}	

\usepackage{xcolor} 

\newcommand{\erosita}{\textit{eROSITA}}

\newcommand{\chandra}{\textit{Chandra}}

\newcommand{\rosat}{\textit{ROSAT}}
\newcommand{\asca}{\textit{ASCA}}
\newcommand{\xmm}{\textit{XMM-Newton}}

\newcommand{\ergps}{erg~cm$^{-2}$~s$^{-1}$}
\newcommand{\aox}{$\alpha_{\rm OX}$}
\newcommand{\edd}{$\lambda_{\rm Edd}$}
\newcommand{\eddxone}{$\lambda_{\rm Edd, X1}$}
\newcommand{\eddxtwo}{$\lambda_{\rm Edd, X2}$}

\let\oldAA\AA
\renewcommand{\AA}{\text{\normalfont\oldAA}}

\title[X-rays from $z=1$ NLS1s]{Narrow-Line Seyfert 1 Galaxies Beyond the Local X-ray Universe: An X-ray spectral sample}

\author[J. Jiang et al.]{Jiachen Jiang$^{1,2}$\thanks{E-mail: Jiachen.Jiang@warwick.ac.uk}, Dominic J. Walton$^{3}$, Luigi C. Gallo$^{4}$, Andrew C. Fabian$^{2}$, Dirk Grupe$^{5}$,\newauthor Richard McMahon$^{2}$, Christopher S. Reynolds$^{6.7}$, Andrew Young$^{8}$, Zhibo Yu$^{9,10}$, \newauthor Honghui Liu$^{11}$, and Zuobin Zhang$^{12}$
\\
$^{1}$Department of Physics, University of Warwick, Gibbet Hill Road, Coventry CV4 7AL, UK\\
$^{2}$Institute of Astronomy, Madingley Road, Cambridge CB3 0HA, UK\\
$^{3}$Centre for Astrophysics Research, School of Physics, Astronomy and Mathematics, University of Hertfordshire, College Lane, Hatfield AL10 9AB, UK\\
$^{4}$Department of Astronomy and Physics, Saint Mary’s University, 923 Robie Street, Halifax, NS, B3H 3C3, Canada\\
$^{5}$Department of Physics, Geology, \& Engineering Technology, Northern Kentucky University, Nunn Drive, Highland Heights, Kentucky 41099, USA\\
$^{6}$Department of Astronomy, University of Maryland, College Park, MD 20742-2421, USA\\
$^{7}$Joint Space-Science Institute, College Park, MD 20742-2421, USA\\
$^{8}$H. H. Wills Physics Laboratory, Tyndall Avenue, Bristol BS8 1TL, UK\\
$^{9}$Department of Astronomy and Astrophysics, The Pennsylvania State University, 525 Davey Lab, University Park, PA 16802, USA\\
$^{10}$Institute for Gravitation and the Cosmos, The Pennsylvania State University, University Park, PA 16802, USA\\
$^{11}$Institut f\"ur Astronomie und Astrophysik, Eberhard-Karls Universit\"at T\"ubingen, D-72076 T\"ubingen, Germany\\
$^{12}$Astrophysics, Department of Physics, University of Oxford, Keble Road, Oxford OX1 3RH, UK
}

\date{Accepted XXX. Received YYY; in original form ZZZ}

\pubyear{2025}

\begin{document}
\label{firstpage}
\pagerange{\pageref{firstpage}--\pageref{lastpage}}
\maketitle

\begin{abstract}
Narrow-line Seyfert 1 AGNs (NLS1s) represent a unique stage in the black hole growth history, characterised by low black hole masses of {approximately $10^{6}$--$10^{8}$ solar masses and around-Eddington} accretion rates. X-ray studies of NLS1s have largely been confined to the local Universe ($z < 0.2$), while their broad-line counterparts and radio-loud quasars have been more extensively investigated at higher redshifts. In this work, we conducted an X-ray spectral analysis for 14 SDSS-observed NLS1s at $z\approx1$ in the eRASS1 catalogue. We found that all of their \erosita\ observations agree with the expected {rest-frame} 2 keV monochromatic luminosity given their {rest-frame} 2500 \AA\ monochromatic luminosity, further supporting evidence of AGN emission. Second, when fitted with a power-law model, most continuum spectra  {between  0.7--7~keV} in their rest frames necessitate photon indices $\Gamma\gtrsim2.5$. Notably, the highest photon index of around 4.7 in one of our NLS1 AGNs hints at a significant contribution from soft excess emission. Finally, our analysis demonstrates that we can align the Eddington ratios with optical measurements by applying {a correction factor between 10-120 to their} X-ray luminosity. Although measurement uncertainty remains considerable, our findings suggest that assumptions for the standard geometrically thin accretion disc model made in previous estimations of this correction factor may not apply to near or super-Eddington NLS1 AGNs. {Finally, we also compare this sample with extremely variable nearby NLS1s and other X-ray-weak AGNs, such as JWST-observed, broad-line AGNs at $z=5-6$, and underscores the importance of deeper X-ray surveys for more X-ray-weak NLS1s.}
\end{abstract}




\section{Introduction}

Narrow-line Seyfert 1 AGNs (NLS1s), frequently distinguished by hosting low-mass black holes (BHs) {approximately $10^{6}-10^{8}$ solar masses with around-Eddington} accretion rates at their centres, exhibit Balmer emission lines with narrower widths {than broad-line Seyfert 1 AGNs}. They often display strong high-ionisation lines typically associated with Seyfert 1 galaxies \citep{davidson78, osterbrock85}. Conventionally, sources are classified as NLS1s if they meet the following criteria: (a) a narrow width of broad Balmer emission lines, with full width at half maximum (FWHM) of H\,${\beta}$ {or Mg\,${\textsc{ii}}\leq$} 2000 km s$^{-1}$; (b) weak [O~\textsc{iii]} forbidden lines; (c) strong Fe~\textsc{ii} emission \citep{osterbrock85, goodrich89, zhou06, rakshit21}.

In the X-ray band, NLS1s frequently manifest rapid and substantial X-ray flux changes, often exhibiting greater amplitudes compared to their broad-line counterparts \citep[e.g.,][]{grupe95, grupe01,fabian09, grupe10, gallo18, alston19, jiang22b}. This flux variability is closely linked to the innermost accretion region of the accretion disc, where the compact X-ray coronal emission originates \citep{boller96, fabian09}. The narrow Balmer lines and rapid fluctuations in X-ray emissions and are believed to result from the relatively small masses of the BHs (approximately $10^{6}-10^{8}$ solar masses) in NLS1s.

The soft X-ray excess represents a prevalent feature observed in numerous NLS1s and continues to be a subject of ongoing research. This is an excess of emission observed when extrapolating the hard X-ray continuum to below 2 keV. The profile of the soft excess emission can be reproduced with a blackbody emission, with temperatures typically ranging from 0.1 to 0.2 keV across various BH mass scales in active galactic nuclei \citep[AGNs, e.g.,][]{gierlinski04}. If this emission comes from the accretion disc, its temperature exceeds what the standard sub-Eddington accretion disc model by \citet{shakura73} can account for. However, it could potentially be elucidated by a slim accretion disc scenario, where photon trapping increases the temperature \citep{abramowicz88}, but only in a super-Eddington accretion regime \citep{tanaka05}.

Two prevailing competing physical models have emerged to explain the origin of the soft excess. One is the warm Comptonisation model, which posits the existence of a warm corona (with temperatures around $kT_\mathrm{e}\sim\ 0.5-1$ keV) that is optically thick ($\tau\sim\ 5-10$) in addition to the hot corona. The soft excess arises from the Comptonisation of UV photons from the disc within this warm corona \citep[e.g.,][]{jin09,petrucci18}. An alternative proposition is the relativistic blurred disc-reflection model, where the emission lines in the soft X-ray band originate from the reflection component and are blurred due to relativistic effects near the BH, thus constituting the soft excess \citep[e.g.,][]{crummy06,walton13, jiang19,jiang20,waddell20}. Support for the reflection model comes from the evidence of soft X-ray reverberation lags \citep[e.g.,][]{fabian09,kara16,chainakun16,demarco19}. The most recent sample-based analysis of bright type-1 AGNs observed by \erosita\ ($F_{\rm 2-10 keV}>10^{-13}$ \ergps) yielded source-dependent conclusions, with some favouring one model over the other \citep{waddell23}.

In-depth modelling and comparison of various spectral models necessitate high signal-to-noise data. NLS1s typically harbour BHs with lower masses compared to their broad-line counterparts. Consequently, for a similar Eddington ratio, NLS1s exhibit lower luminosities than BLS1s and radio-loud quasars, making them more challenging to detect in flux-limited surveys. Hence, many investigations, particularly in the X-ray band, have concentrated on NLS1s within the local Universe, typically at redshifts $z<0.2$. For an example of nearby NLS1 study, the analyses of different local AGN groups have unveiled the widespread presence of the soft excess in both NLS1s and broad-line Seyfert 1 AGNs (BLS1s), with NLS1s {typically} demonstrating a more pronounced soft excess strength \citep[e.g.,][]{puchnarewicz92,grupe04b,middleton07,bianchi09,grupe10,gliozzi20,waddell20}. Moreover, NLS1s often show a softer continuum, characterised by a higher hard X-ray power-law photon index \citep{gliozzi20,waddell20}. {For instance, recent work by \citet{grunwalk23} analysed a sample of approximately 1200 NLS1s observed with \textit{eROSITA} at $z\lessapprox0.8$. By fitting their \textit{eROSITA} spectra with a power-law model, they found a mean photon index of $2.81\pm0.03$. Notably, 10 per cent of the sources exhibited photon indices exceeding 4, indicating an intrinsically very soft X-ray emission.} Independent studies have also identified a positive correlation between the hard X-ray photon index and the Eddington ratio \citep[e.g.,][]{grupe04b,shemmer08,brightman13} as one would expect for NLS1s.

NLS1s may also serve as a crucial stage of low mass and high Eddington ratio {($\lambda_{\rm Edd}=L_{\rm bol}/L_{\rm Edd}$, where $L_{\rm bol}$ is the bolometric luminosity and $L_{\rm Edd}$ is the Eddington luminosity)} in the evolution of SMBHs \citep{grupe99,smita00, grupe04,mathur05}, similar to those in the early Universe \citep[e.g.,][]{maiolino24}. As more data becomes available, it becomes imperative to extend studies to higher redshifts to address questions such as how the accretion physics might differ at various epochs of the Universe compared to our local Universe \citep[e.g.,][]{pacucci24,lambrides24}. 

Previously, we undertook an exploration of five NLS1s {beyond the local Universe, with redshifts} ranging from $z=0.35$ to $z=0.62$, and one at $z=0.92$, utilising archival \textit{XMM-Newton} observations \citep{yu23}. Likely due to selection biases, the BH masses in this sample tend to lie towards the upper end\footnote{The BH masses of the sample in \citet{yu23} were obtained using various methods. A summary of these measurement techniques can be found in Table 1 of \citet{yu23}.} of {the BH mass distribution among local NLS1s}, approximately $m_{\rm BH}=M_{\rm BH}/M_{\odot}=10^{8}$. Through detailed \textit{XMM-Newton} analysis, subsequent to fitting the soft excess model, it was observed that the hard X-ray photon indices were higher than those of local NLS1s at similar Eddington ratios. However, caution was warranted due to uncertainties in Eddington ratio measurements. Nevertheless, the strength of their soft excess emission, a characteristic shared by all objects in the \xmm\ sample, was comparable to that of local NLS1s. Notably, one of the objects, PG 1543$+$489, exhibited {a} relativistic Fe K emission line originating from the inner accretion disc \citep{yu23}.

The German \erosita\ Consortium (eROSITA-DE) has released the first six months of SRG/\erosita\ all-sky survey (eRASS1) data. In this study, we extend our previous work outlined in \citet{yu23} by leveraging these eRASS1 data, thereby advancing our investigation {of NLS1s} to redshifts approximately $z\approx1$. Our focus is directed towards several key questions:

\begin{itemize}
\item Are the detected X-ray luminosities consistent {with expectations} based on the UV luminosity measured by SDSS for typical NLS1s?
\item If so, how do the X-ray and UV luminosities compare to those of local X-ray Universe and quasars at even higher redshifts ($z>4$)?
\item Is the X-ray continuum notably soft? How does it compare to some of the local NLS1s exhibiting potential super-Eddington accretion \citep[e.g.,][]{jiang20,jin23}?
\item Is there discernible evidence of a soft excess? If present, how does the strength of the soft excess compare to that observed in NLS1s within the local Universe? 

{We stress, though, } that due to limited signal-to-noise in the data, our aim is not to differentiate between different models of the soft excess. 
\end{itemize}

\section{\erosita\ Data}

\begin{figure*}
    \centering
    \includegraphics[width=\textwidth]{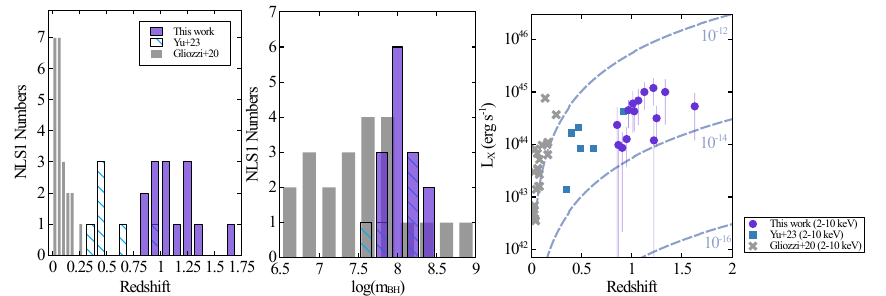}
    \caption{The distribution of source parameters {in the sample of NLS1s discussed here compared to prior samples of such sources presented in the literature}  (left: redshift; middle: BH mass; right: X-ray luminosity vs. redshift). The grey bars show the \xmm-observed NLS1s at low redshift in \citet{gliozzi20}. The blue crossed bars show the \xmm-observed NLS1s between $z=0.35-0.92$ in our previous work \citep{yu23}. The purple bars show the \erosita\ NLS1s at $z>0.85$. The right panel's crosses and squares show the rest-frame 2--10\,keV luminosity of the NLS1s in two samples. The circles show the \erosita\ NLS1s in this work with a detection likelihood larger than 20. {The three dashed curves in the right panel represent constant flux levels of $10^{-12}$, $10^{-14}$ and $10^{-16}$ \ergps, with each point along a given curve corresponding to the same observed flux. } }
    \label{pic_dis}
\end{figure*}

We initiated cross-matching for NLS1s from {the SDSS catalogue \citep{paris18}} of NLS1s at redshifts $z=0.8-2.5$ {as documented in \citet{rakshit21} and the} eRASS1 X-ray point source catalogue \citep{merloni24}. {The high-$z$ NLS1s in \citet{rakshit21} that lack detectable H\,$\beta$ emission within the SDSS wavelength coverage were selected using FWHM(Mg\,\textsc{ii}) < 2000 km s$^{-1}$ as the criterion for classification. This choice was based on the established positive correlation between the line widths of H\,$\beta$ and Mg\,\textsc{ii} \citep{rakshit21}.} Our selection criteria were as follows:
\begin{itemize}
    \item {The angular separation between the eRASS1 X-ray source and SDSS positions, in both RA and Dec, was required to be less than 5.5 arcseconds—approximately one-third of the half-energy width of \erosita's on-axis Point Spread Function \citep{predehl21}—to ensure correct source identification. }
    \item We specifically considered X-ray detections with a {detection} likelihood of 20 for spectral analysis \citep{merloni24}. This criterion was crucial to ensure sufficiently high signal-to-noise ratios for spectral property measurements. 
\end{itemize}

It is important to note that eRASS1 used a methodology that provides independent constraints on the X-ray positional uncertainty using external multi-wavelength source catalogues with known and accurate positions. More details can be found in Section 6.2 in \citet{merloni24}. In particular, \citet{merloni24} used the catalogue of AGNs from \textit{Gaia} and \textit{unWISE} Data \citep{shu19} to cross-match against the eRASS1 X-ray source catalogue. (For readers' interest and future observational reference, we also show the \textit{WISE} and \textit{SDSS} magnitudes of our sample in Table\,\ref{tab_mag}.) We will not repeat the same position uncertainty estimation in this work but adapt the X-ray coordinates presented in the eRASS1 catalogue when selecting sources for our sample. {The 1-$\sigma$ uncertainty in the eRASS1 coordinates (RA and Dec) of our sample ranges from 2 to 5 arcseconds \citep{merloni24}.} {The histogram in Section\,\ref{sec_cor} shows the distribution of coordinate differences (RA and Dec) between SDSS and \erosita.} 


In total, we identified 14 X-ray sources associated with SDSS-observed NLS1s. The information about these NLS1s can be found in Table \ref{tab_src}. The first six columns present the SDSS NLS1 names alongside their corresponding eRASS1 X-ray source names. Additionally, their source positions as measured by SDSS and \textit{eROSITA} are provided. It is noteworthy that in \citet{rakshit21}, the redshifts were spectroscopically measured by SDSS; the rest-frame 3000~\AA\ monochromatic luminosity was used to estimate the bolometric luminosity, applying a global correction factor of 5.15; the BH mass was calculated using the FWHM of {Mg\,\textsc{ii}} \citep{rakshit20}, and the Eddington ratio was derived from the BH mass measurement.

Furthermore, it is essential to acknowledge the selection effects inherent in such flux-limited criteria. We caution that the properties derived from this sample may not necessarily reflect the overall properties of NLS1s at $z\approx1$, but rather are representative of these specific samples. This aspect will be further elucidated and discussed in detail throughout the paper.

Fig. \ref{pic_dis} illustrates the distribution of redshift ($z$) and BH mass ($m_{\rm BH}$) for our \textit{eROSITA} NLS1 sample, juxtaposed with previous \textit{XMM-Newton} samples of NLS1s as documented in \citet{gliozzi20} and \citet{yu23}. \citet{gliozzi20} focused on objects at $z<0.2$, while \citet{yu23} investigated NLS1s spanning from $z=0.35$ to $z=0.62$, with one at $z=0.92$. Notably, our sample exhibits the lowest redshift at $z=0.854$ and the highest value at $z=1.628$. Similar to \citet{yu23}, this flux-limited sample selects BHs with relatively {high} masses compared to the average BH mass {for NLS1s} in the local X-ray Universe ($z<0.2$). The distribution of BH masses in our sample peaks at $10^{8}M_{\odot}$. NLS1s with lower BH masses and luminosities would be fainter, assuming a similar Eddington ratio, and thus less likely to be included in our sample.

In the right panel of Fig. \ref{pic_dis}, we show the distribution of hard X-ray rest-frame 2--10 keV unabsorbed luminosity\footnote{{The rest-frame 2--10 keV unabsorbed flux was calculated after correcting for Galactic column density, using the nominal values from \citet{willingale13}, as listed in Table \ref{tab_pl}. We did not account for additional absorption, such as that from the host galaxy. Due to the limited signal-to-noise ratio of our data, we are unable to constrain any additional modest column density. However, given the soft X-ray nature of our source, significant extra absorption (e.g., above $10^{22}$\,cm$^{-2}$) is not expected. }} versus redshift for the samples in \citet{gliozzi20}, \citet{yu23}, and our work. As a reference, we include constant flux curves for $10^{-16}$, $10^{-14}$, and $10^{-12}$ \ergps\ at various redshifts. The \textit{XMM-Newton} samples in \citet{gliozzi20} primarily probe nearby X-ray bright NLS1s with 2-10 keV fluxes higher than $10^{-12}$ \ergps. In comparison, \citet{yu23} and our study investigate fainter objects with fluxes around or below $10^{-13}$ \ergps.

\begin{figure}
    \centering
    \includegraphics[width=8cm]{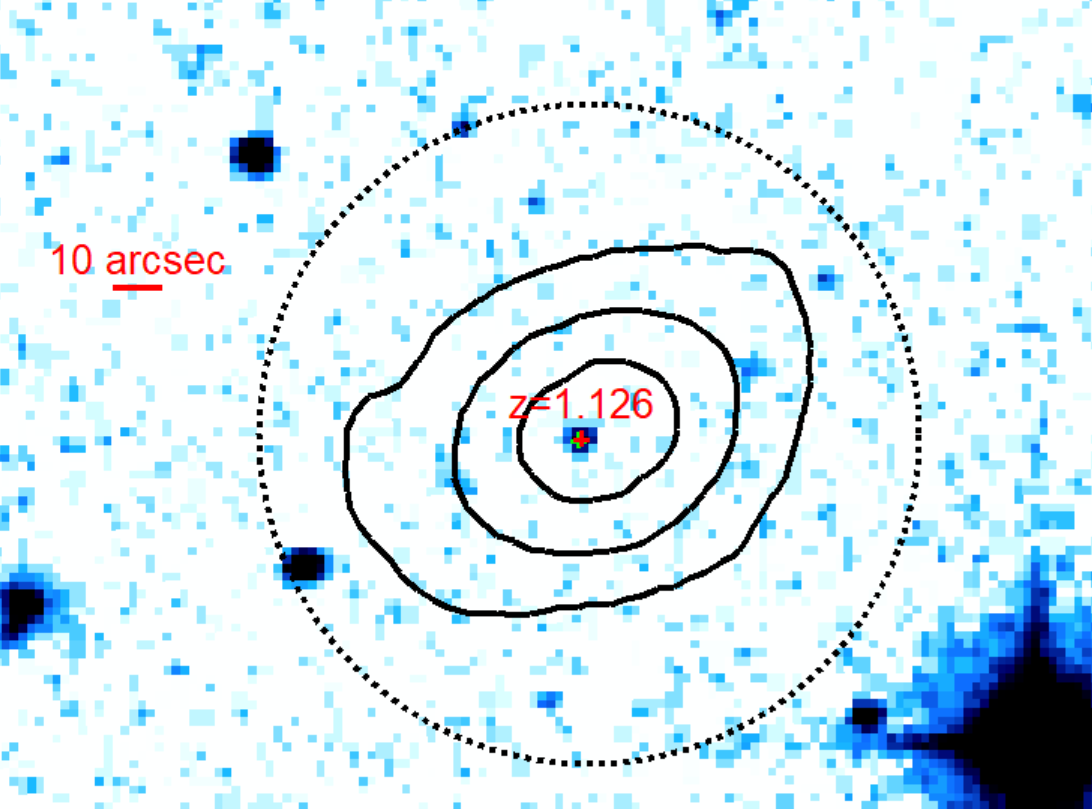}
    \caption{The DSS image of the NLS1 SDSS J094016.02$+$025853.8 ($z$=1.126) overlaid with \erosita\ X-ray contours (three solid {contours representing} 0.01, 0.03 and 0.05 counts per pixel). The dashed circles show the circular region with a radius of 66 arcsec from which the spectrum of this object was extracted. The red cross shows the SDSS position of the galaxy. The green cross shows the position of the X-ray source 1eRASS~J094016.0$+$025853. The two crosses nearly coincide in this image. The reference bar is 10 arcsec in size.}
    \label{pic_src}
\end{figure}

In this work, we used the eRASS1 data products of the 14 $z\approx1$ NLS1s, including spectra, background spectra, response matrices, and ancillary response files, which were extracted using SRCTOOL \citep{brunner22}. The tool selected a circular source extraction region to optimise the signal-to-noise ratio of the source spectrum, considering the local background surface brightness and the shape of the point spread function (PSF). The radius can not be less than 15 arcsec or higher than 99\% of the PSF encircled energy fraction, assuming a circular PSF, and takes into account any excluded neighbouring contaminating sources. The circular regions were centred at the positions of the X-ray sources rather than the SDSS positions. We emphasise that due to our strict criteria requiring a RA/DEC coordinate difference smaller than 5.5 arcsec, we would not anticipate significant changes in the PSF if one were to use SDSS positions to extract data products. The background regions were configured as annulus regions. The radii of the circular source and background annulus regions for each object can be found in Table \ref{tab_regions}. {The positions of contaminating sources excluded in the background regions can be accessed via the eRASS1 website\footnote{https://erosita.mpe.mpg.de/dr1/erodat/catalogue/search.}.} Figure \ref{pic_src} illustrates an example of the source region alongside the \erosita\ X-ray contours for J0940.

The luminosity distances in this study were computed based on the cosmological constants outlined in \citet{planck20} {with a flat $\Lambda$CDM cosmology: Hubble constant of $H_{0}=67.36$~km~s$^{-1}$~Mpc$^{-1}$, matter density of $\Omega_{m}=0.315$, and an effective mass density of dark energy of $\Omega_{\Lambda}=0.685$.} We grouped the spectra to ensure a minimum of 2 counts per energy bin using GRPPHA. For spectral analysis, we employed XSPEC \citep{arnaud96}, utilising C statistics as a measure of goodness of fit \citep{cash79}. {Section \ref{sec_pl} shows our spectral analysis by directly subtracting their background spectra. Additionally, we considered modelling the background spectra and applied the best-fit background models when fitting the source-region spectra in Section\,\ref{sec_bkg_model}.}

\begin{table*}
    \centering
        \caption{The names and positions of the SDSS galaxies and their corresponding \erosita\ X-ray sources identified in eRASS1. The last four columns were optical SDSS measurements in \citet{rakshit21}.  In particular, the Eddington ratios $\lambda_{\rm Edd}$ were derived from the 3000 \AA\ monochromatic luminosity in \citet{rakshit21}. In this work, we refer to the galaxies of our sample as the short names. $^{*}$The unabsorbed monochromatic luminosity at rest-frame 3000 \AA\ in units of erg s$^{-1}$.}
    \label{tab_src}
    \begin{tabular}{ccccccccccc}
\hline\hline
SDSS {Name} & RA & Dec & eRASS {Name} & RA & Dec & Short {Name} & $z$ & $\log(\lambda L_{\lambda})^{*}$ & $\log(m_{\rm BH})$ & $\log(\lambda_{\rm Edd})$ \\
\hline
023532.20-083052.7 & 38.88420 & -8.51463 & 023532.1-083054 & 38.88407 & -8.51509 & J0235 & 1.628 & 45.71 & $8.3 \pm 0.3$ 	 & -0.03\\
082455.47+391641.8 & 126.23117 & 39.27831 & 082455.6+391639 & 126.23177 & 39.27776 & J0824 & 1.216 & 45.82 & $8.2 \pm 0.2$ 	 & 0.18\\
082604.55+294212.6 & 126.51900 & 29.70352 & 082604.5+294210 & 126.51893 & 29.70279 & J0826 & 1.065 & 45.52 & $7.7 \pm 0.3$ 	 & 0.29\\
084508.99+173518.0 & 131.28748 & 17.58834 & 084508.9+173519 & 131.28723 & 17.58882 & J0845 & 0.905 & 45.63 & $8.2 \pm 0.2$ 	 & -0.00\\
085925.04+215620.0 & 134.85436 & 21.93893 & 085925.0+215620 & 134.85455 & 21.93914 & J0859 & 0.968 & 45.28 & $7.9 \pm 0.4$ 	 & -0.14\\
092149.48+082646.8 & 140.45621 & 8.44635 & 092149.5+082645 & 140.45635 & 8.44589 & J0921 & 0.867 & 45.29 & $8.06 \pm 0.08$ 	 & -0.20\\
094016.02+025853.8 & 145.06676 & 2.98162 & 094016.0+025853 & 145.06692 & 2.98153 & J0940 & 1.126 & 45.89 & $8.4 \pm 0.1$ 	 & 0.09\\
095748.05+070440.5 & 149.45026 & 7.07792 & 095747.9+070439 & 149.44988&7.07750 & J0957 & 1.025 & 45.33 & $8.0 \pm 0.3$ 	 & -0.10\\
102634.33+320135.0 & 156.64304 & 32.02640 & 102634.3+320136 & 156.64298&32.02684 & J1026 & 1.221 & 45.17 & $8.0 \pm 0.1$ 	 & -0.29\\
103636.21+240551.7 & 159.15092 & 24.09772 & 103636.3+240554 & 159.15142&24.09836 & J1036 & 1.011 & 45.49 & $7.9 \pm 0.2$ 	 & 0.19\\
104537.54+010337.6 & 161.40641 & 1.06046 & 104537.2+010339 & 161.40534&1.06111 & J1045 & 1.333 & 45.54 & $8.1 \pm 0.2$ 	 & 0.01\\
113355.79-012913.7 & 173.48251 & -1.48720 & 113355.5-012914 & 173.48164&-1.48743 & J1133 & 1.248 & 45.78 & $8.27 \pm 0.02$ 	 & 0.08\\
114852.67+245715.8 & 177.21949 & 24.95439 & 114852.9+245717 & 177.22073&24.95494 & J1148 & 0.854 & 44.74 & $7.82 \pm 0.05$ 	 & -0.51\\
130405.34-013904.2 & 196.02228 & -1.65120 & 130405.3-013900 & 196.02214&-1.65011 & J1304 & 0.950 & 45.26 & $8.0 \pm 0.1$ 	 & -0.15\\
\hline\hline
    \end{tabular}
\end{table*}

\section{X-ray Properties of \erosita\ $z\approx1$ NLS1s}

In this section, we investigate the X-ray and UV luminosity of these NLS1s. Our focus centers on the inquiry of whether the X-ray luminosity of the associated X-ray sources of the SDSS-observed NLS1s aligns with our expectations given the UV luminosity. Specifically, we compute \aox, a metric describing the ratio of UV and X-ray monochromatic luminosity.

\subsection{UV and X-ray monochromatic luminosity}

We first computed the rest-frame 2500 \AA\ monochromatic luminosity for the NLS1s in our sample. While \citet{rakshit21} did not provide the 2500 \AA\ monochromatic luminosity for the SDSS samples, we were able to derive the AGN luminosity by using the best-fit SDSS spectral models for the host galaxy-subtracted spectra in \citet{rakshit21}. The AGN continuum (flux density) was characterised by a power law $f_{\lambda}=\beta(\lambda/\lambda_{0})^{\alpha}$, where the reference wavelength $\lambda_{0}=3000\AA$. Here, $\beta$ and $\alpha$ represent the normalisation of the power law in units of erg cm$^{-2}$ s$^{-1}$ \AA$^{-1}$ and the dimensionless power-law index, respectively. From this, we calculated the monochromatic luminosity of our NLS1s at 2500 \AA\ using the flux density, assuming isotropic emission.

Next, we used \textit{eROSITA} data to compute the rest-frame 2 keV monochromatic luminosity by measuring the unabsorbed 2 keV flux density and similarly assuming isotropic X-ray emission. These results are presented in Table \ref{tab_flux}. Based on the 2 keV and 2500 \AA\ monochromatic luminosity, we further calculated \aox\ using the formula \aox=$0.3838\log(L_{\rm 2keV}/L_{2500 \AA})$ \citep[e.g.,][]{strateva05}. The values of \aox\ are also reported in Table \ref{tab_flux}.

\begin{table}
    \centering
    \caption{The monochromatic luminosity at 2 keV and 2500 \AA\ in units of erg s$^{-1}$ Hz$^{-1}$ and corresponding \aox\ of the NLS1s in this sample.}
    \label{tab_flux}
    \begin{tabular}{cccc}
    \hline\hline
    Names & $\log(L_{\rm 2keV})$ & $\log(L_{2500 \AA})$ & \aox\\
          & erg s$^{-1}$ Hz$^{-1}$ & erg s$^{-1}$ Hz$^{-1}$ \\
    \hline
J0235 & $26.8\pm0.3$ & $30.480\pm0.002$ & $-1.43\pm0.11$ \\
J0824 & $26.8\pm0.2$ & $30.296\pm0.005$ & $-1.35\pm0.09$ \\
J0826 & $26.5\pm0.3$ & $30.607\pm0.001$ & $-1.56\pm0.10$ \\
J0845 & $26.6\pm0.3$ & $30.855\pm0.001$ & $-1.65\pm0.10$ \\
J0859 & $27.0\pm0.2$ & $30.265\pm0.002$ & $-1.25\pm0.09$ \\
J0921 & $26.5\pm0.3$ & $30.480\pm0.002$ & $-1.51\pm0.10$ \\
J0940 & $26.8\pm0.2$ & $30.265\pm0.002$ & $-1.35\pm0.09$ \\
J0957 & $26.9\pm0.2$ & $30.607\pm0.001$ & $-1.44\pm0.10$ \\
J1026 & $26.5\pm0.3$ & $30.607\pm0.001$ & $-1.59\pm0.11$ \\
J1036 & $26.5\pm0.3$ & $30.607\pm0.001$ & $-1.56\pm0.10$ \\
J1045 & $27.0\pm0.2$ & $30.480\pm0.002$ & $-1.32\pm0.10$ \\
J1133 & $27.0\pm0.3$ & $30.607\pm0.001$ & $-1.38\pm0.10$ \\
J1148 & $26.5\pm0.3$ & $30.480\pm0.002$ & $-1.52\pm0.10$ \\
J1304 & $26.3\pm0.3$ & $30.332\pm0.005$ & $-1.54\pm0.10$ \\
\hline\hline
    \end{tabular}
\end{table}

\begin{figure*}
    \centering
    \includegraphics[width=\textwidth]{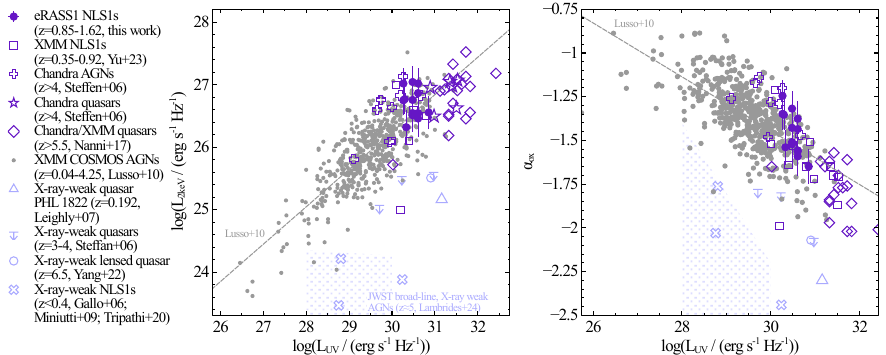}
    \caption{Left: X-ray (2 keV) vs UV (2500 \AA) monochromatic luminosity of the NLS1s at $z\approx1$ (purple circles) in comparison with other samples of AGNs. Grey circles: \xmm\ COSMOS survey of type-1 AGNs at $z=0.02-4.25$ \citep{lusso10}. The dashed straight line is the X-ray--UV luminosity positive correlation derived from the COSMOS survey \citep{lusso10}. Open squares: $z=0.35-0.92$ NLS1s \citep{yu23}; open purple stars and crosses: $z>4$ \chandra-observed quasars and radio-quiet AGNs \citep{steffen06}; open diamonds: $z>5.5$ \xmm\ or \chandra-observed quasars \citep{nanni17}. {The light blue symbols represent the monochromatic luminosity of different AGN populations with weak X-ray emission. The open triangle: the nearby X-ray-weak quasar PHL 1822 at $z=0.192$ \citep{leighly07}; light blue arrows: the upper limits of the 2 keV monochromatic luminosity of $z\approx3-4$ X-ray-weak quasars in the COMBO-17 and extended \chandra\ Deep Field-South Surveys \citep{steffen06}; the open light blue circle: an X-ray-weak, lensed quasar at $z=6.5$ \citep{yang22};  diagonal open crosses: three NLS1s in their X-ray-weak states (1H 0707$-$495, Mrk 335 and PHL 1092) in the nearby Universe \citep{gallo06,miniutti09,tripathi20}; the shaded region: the estimated 2 keV and 2500 \AA\ monochromatic luminosity of X-ray-weak, JWST-observed, and broad-line AGNs modified from Fig. 1 in \citet{lambrides24}. The upper edge of the shaded region shows the upper limit of their 2 keV monochromatic luminosity.  Right: \aox\ vs UV (2500 \AA) monochromatic luminosity for the same populations.}}
    \label{pic_xuv}
\end{figure*}

We compared the rest-frame 2500 \AA\ and 2 keV monochromatic luminosity in our sample of NLS1s with those of \textit{XMM-Newton}-observed AGNs in the COSMOS Survey as presented in \citet{lusso10}. The COSMOS Survey in \citet{lusso10} comprised 545 radio-quiet X-ray-selected type-1 AGNs in the redshift range of $z=0.04-4.25$, including 322 spectroscopically selected type-1 and 233 SED-selected type-1 AGNs. The X-ray monochromatic luminosity and UV monochromatic luminosity of our sample of NLS1s at $z\approx1$ closely align with the correlation between these parameters derived by \citet{lusso10}, as shown by the dashed line in Fig. \ref{pic_xuv}.

\subsection{Similar to the `simple' X-ray NLS1s in the local Universe}

\citet{gallo06} previously investigated a sample of nearby NLS1s ($z<0.2$). Based on the spectral complexity, such as whether the rest-frame 2.5-10 keV spectra are consistent with an absorbed power law plus a narrow Gaussian line profile for the Fe K emission, \citet{gallo06} categorised the NLS1s into two groups: `simple' and `complex'. The `simple' NLS1s all exhibit \aox\ that is consistent with what one would expect using a typical $L_{2500\AA}$ and \aox\ relationship \citep{strateva05}. On the other hand, the `complex' NLS1s often display X-ray weakness, such as 1H 0707$-$495, one of the most variable X-ray NLS1s \citep{fabian09,dauser12,boller21}. NLS1s can also transition between these two groups when the X-ray luminosity undergoes significant changes. During the X-ray weak stage, their X-ray spectrum often reveals strong soft excess emission and significant relativistic accretion disc spectra \citep[e.g.,][]{parker14,jiang18}.

Our sample of NLS1s at $z\approx1$ exhibits an X-ray monochromatic luminosity that is consistent with the expected values given the 2500 \AA\ monochromatic luminosity, indicating that they are mostly similar to the `simple' NLS1s studied in the local X-ray Universe. 

\subsection{Compared to X-ray-weak AGNs} \label{sec_uvx}

{X-ray weak AGNs are particularly intriguing, as they may signify a distinct phase of potentially super-Eddington BH growth \citep[e.g.,][]{wang14,pacucci24,inayoshi24}. Additionally, they highlight the necessity of complementary multi-wavelength coverage in AGN surveys \citep[e.g.,][]{barchiesi21,cappelluti24}.}

{One particularly noteworthy NLS1, WISEA~J033429.44$+$000610.9 (J0334) at $z=0.35$, investigated in our previous work \citep{yu23}, exhibited exceptionally weak X-ray emission while displaying the most pronounced soft excess among the \xmm\ sample in \citet{yu23}. Its characteristics are more similar to those of `complex' NLS1s, as indicated by the bottom blue square in Fig. \ref{pic_xuv}. In this eRASS sample, the object with the lowest \aox\ is J0845; however, its value remains consistent with the expected correlation from \citet{lusso10} within measurement uncertainties. J0845 does not exhibit an X-ray state comparable to that of J0334 or certain other local NLS1s \citep[e.g.,][]{parker14}. }

{We compared the combined sample of X-ray NLS1s from this work and \citet{yu23} with several known Seyfert AGNs exhibiting weak X-ray emission, represented by the light blue symbols in Fig.\,\ref{pic_xuv}. As previously noted, the `complex' NLS1s often display large-amplitude X-ray variability and enter a low X-ray flux state. Examples include Mrk~335, which has been observed with \aox=$-2.03$ \citep[e.g.,][]{tripathi20}, 1H~0707$-$495 with \aox=$-1.76$ \citep{gallo06}, and PHL1092 with \aox=$-2.44$ \citep{miniutti09}. Other extreme cases include RX~J0134$-$4258, where \aox\ declined from $-1.47$ to $-2.00$ over 1–2 years captured \rosat\ and \asca\ \citep{grupe00}, and the bare Seyfert 2 AGN 1ES~1927$+$654, whose \aox\ varied from -0.82 to -1.82, or even lower within months \citep{gallo13, laha22}.}

{Weak X-ray emission is not exclusive to Seyfert AGNs; it is also observed in quasars. In Fig.\,\ref{pic_xuv}, we include several X-ray weak quasars spanning a wide redshift range, from the nearby Universe \citep[e.g., PHL~1822 at $z=0.192$;][]{leighly07}, to $z=3-4$ in the \chandra\ Deep Field-South Survey \citep[][where only upper limits on the 2 keV monochromatic luminosity were obtained]{steffen06}, and even to the distant Universe \citep[e.g., an X-ray-weak lensed quasar at $z=6.5$;][]{yang22}. These X-ray-weak quasars all exhibit an \aox\ of less than approximately -1.8. The population of X-ray-weak AGNs is growing rapidly. For brevity, we only show these objects in Fig.\,\ref{pic_xuv} as an example.}

{Among these X-ray-weak AGNs, a particularly intriguing population warrants mention—compact galaxies characterised by a `v-shaped' spectral energy distribution, featuring a blue UV continuum below approximately rest-frame 1000–2000 \AA\ and a red optical continuum at longer wavelengths \citep[e.g.,][]{furtak23a,setton24,labbe25}. JWST spectra have revealed their broad H$\alpha$ and H$\beta$ emission, indicative of a rapidly accreting SMBH \citep[e.g.,][]{fujimoto23c,killi24,kocevski24,kokorev24,ubler23,matthee24}. Notably, these objects exhibit extremely weak X-ray emission \citep[e.g.,][]{yue24}. Their estimated location in the $L_{2500 \AA}$ versus \aox\ parameter space is shown as the shaded region in Fig.\,\ref{pic_xuv}. The upper boundary of this region represents the upper limit of their 2 keV monochromatic luminosity, as estimated by \citet[][see their Fig. 1 for individual source values\footnote{Note that the 2500 \AA\ monochromatic luminosity in \citet{lambrides24} was derived from the line width of the H$\alpha$ emission lines.}]{lambrides24}.}

{The weak X-ray emission observed in these special AGNs, both at low and high redshifts, is likely attributed to various factors or a combination of multiple mechanisms, \textit{depending on individual targets}. Explanations include extreme intrinsic X-ray variability of the primary coronal emission \citep[e.g,][]{dauser12,parker14,jiang18}, super-Eddington slim accretion disc \citep[e.g.,][]{pacucci24}, or transient events such as the disruption of the X-ray corona by a tidally disrupted star \citep[e.g.,][]{ricci21} or a magnetic field reset \citep{scepi21}. Additionally, complex and variable absorption features have been identified in some weak X-ray states \citep[e.g.,][]{boller21,liu21,yang22}. We will further discuss the implications and significance of identifying more X-ray-weak AGNs in Section\,\ref{sec_xrayweaknls1}.}

\subsection{Optical Eddington ratio and \aox}

\begin{figure}
    \centering
    \includegraphics[width=8cm]{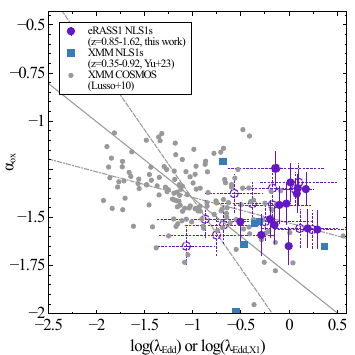}
    \caption{Eddington ratio $\lambda_{\rm Edd}$ vs \aox\ for $z\approx1$ NLS1s (purple circles), $z=0.35-0.92$ NLS1s \citep[blue squares][]{yu23} and the \xmm\ COSMOS AGNs \citep[grey circles][]{lusso10}. The three dashed lines show the correlation between two parameters derived from the COSMOS survey using three different methods in \citet{lusso10}: {linear regression OLS(\aox|\edd) treating \edd\ as the independent variable (the dashed line in the figure), OLS(\edd|\aox) treating \aox\ as the independent variable (the dash-dotted line), and the bisector of the two regression lines (the solid line)}. The filled purple circles use the Eddington ratios $\lambda_{\rm Edd}$ calculated in \citet{rakshit21} based on 3000 \AA\ luminosity. The open purple circles use the Eddington ratios $\lambda_{\rm Edd, X1}$ calculated in this work using 2--10 keV unabsorbed luminosity and an X-ray correction factor of 20.}
    \label{pic_alpha}
\end{figure}

\aox\ has been found to exhibit a correlation with the Eddington ratio $\lambda_{\rm Edd}$ of the accretion process. For instance, \citet[][]{lusso10} examined type-1 AGNs in the \xmm\ COSMOS Survey and identified a negative correlation between \aox\ and \edd\ among the samples of type-1 AGNs, as depicted by the grey circles {in Fig.}\,\ref{pic_alpha}. Depending on the statistical methods employed, three {descriptions of the statistical correlation} arise \citep{lusso10}: Ordinary Least Square regression OLS(\aox|\edd) (treating \edd\ as an independent parameter, shown by the dashed line in Fig. \ref{pic_alpha}), OLS(\edd|\aox) (treating \aox\ as an independent parameter, shown by the dash-dotted line in Fig. \ref{pic_alpha}), and the bisector of the two regression lines (solid line in Fig. \ref{pic_alpha}). We overlaid the NLS1s from our previous work \citep{yu23} and the present study onto the same diagram. The eRASS $z\approx1$ NLS1s are depicted by the filled purple circles. In this section, we specifically focus on the optical Eddington ratio. These ratios are calculated using the 3000 \AA\ AGN intrinsic luminosity, corrected by a bolometric correction factor of 5.15 \citep{rakshit21}. The corresponding values are provided in Table \ref{tab_src}.

Comparing with the \xmm\ COSMOS Survey samples, our sample of $z\approx1$ NLS1s {may exhibit a marginally } higher \aox\ at a similar Eddington ratio. With future deeper X-ray surveys, we anticipate discovering more NLS1s in the distant Universe with lower X-ray luminosity or lower \aox\ to further statistically determine the \aox-\edd\ correlation at the higher end of \edd\ distribution.

Meanwhile, we also raise caution regarding the systematic uncertainty in estimating \edd. In {Section \ref{sec_xedd}}, we will also present X-ray Eddington ratios, \eddxone\ and \eddxtwo, by applying correcting factors to the 2-10 keV intrinsic luminosity. We argue that by employing this approach, we systematically decrease the estimation of \edd, and our samples of NLS1s align better with the correlation observed in the \xmm\ COSMOS Survey. See the open circles in Fig. \ref{pic_alpha} where a constant X-ray bolometric luminosity correction factor was applied to calculate Eddington ratios.

\section{X-ray Spectral Fitting} \label{sec_pl}

\subsection{Modelling their very soft X-ray continuum emission}

\begin{figure*}
    \centering
    \includegraphics[width=\textwidth]{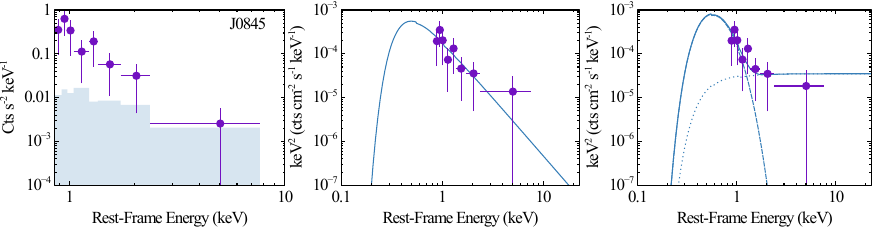}
    \includegraphics[width=\textwidth]{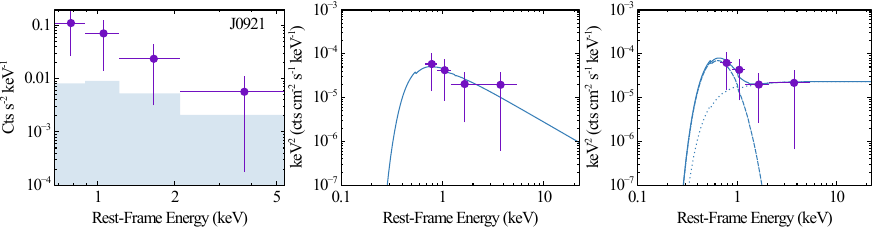}
    \includegraphics[width=\textwidth]{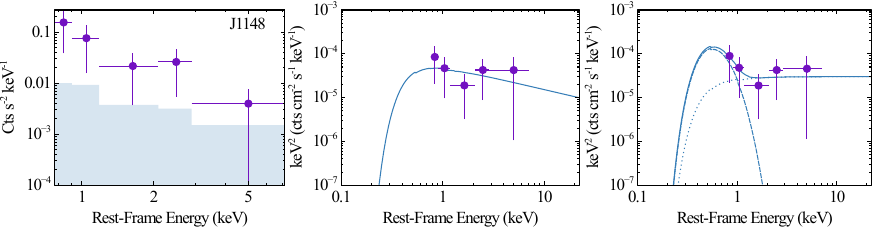}
    \caption{Three examples of \erosita\ X-ray spectra of the NLS1s in our sample. Left: folded count rate spectrum (purple crosses) and background spectrum (shaded regions); middle: unfolded spectrum and the best-fit power-law model; right: unfolded spectrum and the best-fit model including a blackbody component accounting for the soft excess emission. See Section\,\ref{sec_pl} for best-fit power-law models and Section\,\ref{sec_soft} for more detailed soft excess modelling.}
    \label{pic_model}
\end{figure*}

We fitted the \textit{eROSITA} spectra of the NLS1s using a redshifted power-law model (\texttt{zpowerlw}) in XSPEC. The redshift parameters were fixed at the optical {spectroscopic} redshifts. The foreground Galactic column density was fixed at the nominal values calculated in \citet{willingale13} and listed in Table \ref{tab_pl}. Most data could be adequately described with such a power-law model. Table \ref{tab_pl} presents the C-stat and best-fit photon indices. Three examples of the best-fit models and corresponding spectra for J0845, J0921, and J1148 are illustrated in Fig. \ref{pic_model}. Additional spectral fitting results can be found in Figs. \ref{pic_ap1}, \ref{pic_ap2}, and \ref{pic_ap3} in the Appendix. We also considered modelling the background spectrum and applied the best-fit background model when fitting the source-region spectrum in Section\,\ref{sec_bkg_model} rather than directly subtracting the background spectrum. By doing so, we found a consistent photon index measurement.

Our spectra exhibit very high photon indices, indicating a remarkably soft continuum emission from these NLS1s. It is important to note that, due to the limited X-ray flux of our targets, {the current \erosita\ data for these sources only} probe up to approximately 7 keV in the rest frame, beyond which the targets fall below the current sensitivity. Among them, J0826, J1036, and J1045 demonstrate the lowest photon index, around 2. The highest photon index is observed in J0845, which is $4.7^{+0.9}_{-1.2}$, significantly higher than the highest values of the local NLS1s' X-ray spectral index, around 2.8-2.9 during the X-ray luminous state  \citep[e.g.,][]{jiang18}. Considering the good agreement between the UV and X-ray luminosity presented in the previous section and the very soft continuum emission, we propose the following interpretations: first, the X-ray emission does originate from the innermost accretion region, similar to most other local NLS1s. Second, the very soft X-ray continuum emission likely arises from a combination of hot coronal and soft excess emissions.

\begin{figure}
    \centering
    \includegraphics[width=8cm]{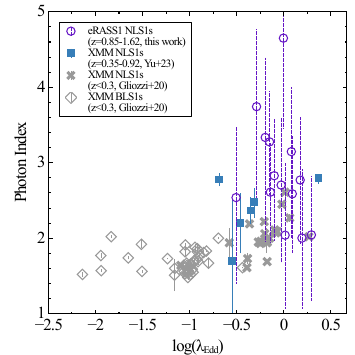}
    \caption{Best-fit X-ray photon indexes for the full band \erosita\ spectra vs Eddington ratios calculated using optical 3000 \AA\ luminosity (open purple circles). Blue squares: $z=0.35-0.92$ NLS1s \citep{yu23}; grey crosses: nearby NLS1s; grey diamonds: nearby BLS1s \citep{gliozzi20}. All NLS1s show a very soft continuum with a photon index around or higher than 2.5. The average value {for the NLS1 sample discussed here} is higher than the nearby NLS1s at $z<0.3$ \citep{gliozzi20}, indicating the possibility of significant soft excess emission. Note that the other photon index measurements in \citet{gliozzi20,yu23} were obtained by modelling the soft excess with an additional blackbody model. }
    \label{pic_gamma}
\end{figure}

We compare our best-fit photon index with the measurements of NLS1s in \citet{gliozzi20} and \citet{yu23}, as well as BLS1s in \citet{gliozzi20}, in Fig. \ref{pic_gamma}. \citet{gliozzi20} highlighted the significantly softer X-ray continuum in NLS1s compared to BLS1s, which aligns with the expectation for a higher Eddington ratio in NLS1s than in BLS1s. The $z\approx1$ NLS1s in our \textit{eROSITA} sample exhibit an even softer continuum with a median value of 2.7, {which is consistent with findings from a much larger sample of approximately 1200 NLS1s observed in eRASS1 at $z\lessapprox0.8$ by \citet{grunwalk23}, who reported a median photon index of $2.81\pm0.03$ based on full-band spectral fits using a single power-law model.}

We exercise caution in interpreting these results due to large measurement uncertainty, although we are able to rule out the possibility of a lower limit at $\Gamma=2.5$ with 90\% confidence in a few cases, such as J0845. The photon index measurements in \citet{gliozzi20} and \citet{yu23} were also obtained after carefully modelling the soft excess, for instance, with a blackbody model. In contrast, we measured the photon index by fitting the full band data between rest-frame 0.7-7 keV. As concluded earlier, a significant contribution of the soft excess emission explains the very soft X-ray continuum. 

\begin{table}
    \centering
        \caption{Best-fit power-law photon indices for the full band spectra. The Galactic column density $N_{\rm H}$ is fixed at nominal values in \citet{willingale13}. $\nu$ is the degree of freedom.}
    \label{tab_pl}
    \begin{tabular}{cccc}
    \hline\hline
     Names    & $N_{\rm H}$ & Photon Index & C-stat/$\nu$\\
              & $10^{20}$ cm$^{-2}$ \\
     \hline
J0235 & $3.6$ & $2.7^{+0.9}_{-1.0}$ & 1.96/3 \\
J0824 & $4.4$ & $2.8^{+0.8}_{-0.8}$ & 4.82/6 \\
J0826 & $3.9$ & $2.0^{+0.8}_{-0.9}$ & 5.97/3 \\
J0845 & $2.2$ & $4.7^{+0.9}_{-1.2}$ & 1.88/6 \\
J0859 & $3.3$ & $2.6^{+0.6}_{-0.7}$ & 9.12/6 \\
J0921 & $4.2$ & $3.3^{+1.1}_{-1.4}$ & 0.34/2 \\
J0940 & $3.5$ & $2.6^{+0.8}_{-0.8}$ & 4.24/5 \\
J0957 & $2.9$ & $2.8^{+0.8}_{-0.9}$ & 4.99/4 \\
J1026 & $2.2$ & $3.7^{+1.0}_{-1.2}$ & 6.51/2 \\
J1036 & $2.2$ & $2.0^{+0.9}_{-0.9}$ & 6.06/3 \\
J1045 & $4.1$ & $2.0^{+1.0}_{-1.0}$ & 0.96/3 \\
J1133 & $2.5$ & $3.1^{+0.8}_{-1.1}$ & 2.92/3 \\
J1148 & $2.2$ & $2.5^{+0.9}_{-1.1}$ & 1.86/3 \\
J1304 & $1.7$ & $3.3^{+0.7}_{-0.7}$ & 4.65/6 \\
\hline\hline
    \end{tabular}
\end{table}

\subsection{An attempt to constrain the soft excess} \label{bbfit}

In this section, we will dissect the very soft X-ray continuum emission of these \textit{eROSITA}-observed NLS1s into two components: hard X-ray Comptonisation from the hot coronal region and the soft excess emission. Due to the limited signal-to-noise ratio, our goal is not to distinguish between different models for the soft excess emission, such as disc reflection or warm corona, by exhaustively exploring all possibilities as in \citet{jiang22c}. Instead, we adopt the simplest blackbody model, following the approach in \citet{gliozzi20} and \citet{yu23}, and focus on the strength of the soft excess. We apply the \texttt{zashift} model to \texttt{bbody} in XSPEC to account for the sources' redshifts.

Due to the limited signal-to-noise, we needed to fix the temperature of the \texttt{bbody} component at 0.1 keV and the photon index of the hard X-ray power law at 2, which is the median value of the hard X-ray photon index of local NLS1s (see the grey crosses in Fig. \ref{pic_gamma}). This approach assumes a global AGN X-ray spectral template but allows each component's strength to be variable in our spectral fitting. 

In three cases—J0826, J1036, and J1045—the normalisation of the soft excess emission calculated by the blackbody component is completely unconstrained. In these instances, we find that their \textit{eROSITA} spectra are consistent with the hardest continuum in our sample, with $\Gamma\approx2$. 

The best-fit flux parameters of each component are shown in Table \ref{tab_bb}. Please refer to Fig. \ref{pic_ap1}, \ref{pic_ap2}, and \ref{pic_ap3} for their best-fit models. 

\section{Hard X-ray luminosity and Eddington ratios} \label{sec_xedd}

\begin{figure*}
    \centering
    \includegraphics[width=\textwidth]{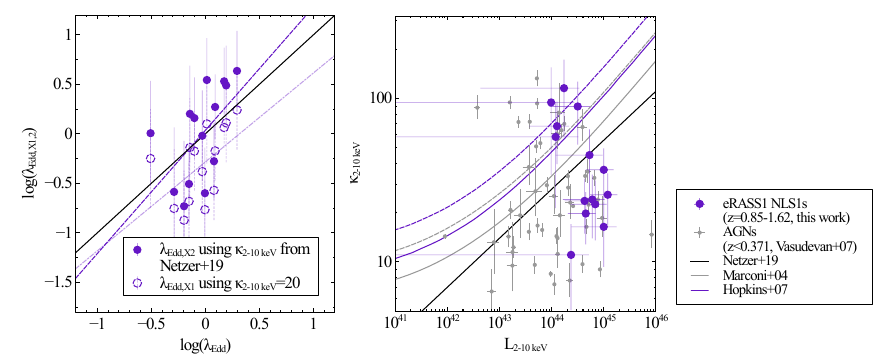}
    \caption{Left: Eddington ratios calculated based on 2--10 keV unabsorbed luminosity (this work) vs optical 3000 \AA\ luminosity \citep{rakshit21}. Open circles represent $\lambda_{\rm Edd, X1}$ using a global X-ray correction factor of 20. Filled circles represent $\lambda_{\rm Edd, X2}$ using the X-ray correction factors derived by \citet{netzer19}, which depends on X-ray luminosity. The dashed (dash-dotted) line in the left panel of Fig. \ref{pic_kappa} represents the best-fit correlation between $\lambda_{\rm Edd,X2}$ (\eddxone) and $\lambda_{\rm Edd}$.  Right: X-ray correction factor $\kappa$ derived by assuming the precision of the Eddington ratio calculated using optical 3000 \AA\ luminosity. The solid black, grey and purple lines show the theoratical X-ray correction factors as a function of X-ray luminosity in \citet{netzer19}, \citet{marconi04} and \citet{hopkins07}, respectively. The dashed grey and purple lines would be the same lines from \citet{marconi04} and \citet{hopkins07} if IR emission were included at the level of one third of the total luminosity \citep{vasudevan07}. The grey circles show the observational measurements of $\kappa$ in the sample of AGNs at $z<0.37$ in \citet{vasudevan07}.}
    \label{pic_kappa}
\end{figure*}

\subsection{X-ray Eddington ratios}

Using the best-fit models, we estimated the unabsorbed flux in the rest-frame 2--10 keV band, which was then used to calculate the Eddington ratio. For J0826, J1036, and J1045, where the spectra are consistent with a simple power law of $\Gamma = 2$, we based our flux estimates on the best-fit power law model. For the remaining objects in the sample, the best-fit photon index exceeds $\Gamma = 2$. Although adding a blackbody component to account for soft excess does not statistically improve the fit for these cases, we included it in our calculation. This approach assumes $\Gamma = 2$ and helps avoid underestimating the 2--10 keV luminosity by considering a softer continuum.

We then calculated the X-ray Eddington ratios in the following two ways. First, we assume a global bolometric luminosity correction factor of 20, denoted as $\kappa=L_{\rm bol}/L_{\rm 2-10 keV}=20$. This choice is based on the average values of local AGNs \citep[e.g., with $L_{\rm bol}\lesssim10^{46}$ erg s$^{-1}$, ][]{duras20}. Note that $10^{46}$ erg s$^{-1}$ is around the Eddington luminosity for a $m_{\rm BH}=10^{8}$ BH, similar to the ones in our sample. The notation of this X-ray Eddington ratio is $\lambda_{\rm Edd,X1}$.

Second, to be consistent with the estimation in our previous work \citet{yu23}, we also adopted a luminosity-dependent bolometric correction factor as in \citet{netzer19}, $\kappa=7\times(L_{\rm 2-10 keV}/10^{42})^{0.3}$, where $L_{\rm 2-10 keV}$ is in units of erg s$^{-1}$. The notation of this X-ray Eddington ratio is  $\lambda_{\rm Edd,X2}$. {The values of $\lambda_{\rm Edd,X1}$ and $\lambda_{\rm Edd,X2}$ are shown in Table\,\ref{tab_eddington}.}

We compared the calculated X-ray Eddington ratios, $\lambda_{\rm Edd,X1}$ and $\lambda_{\rm Edd,X2}$, with the optical Eddington ratios calculated using the 3000~\AA\ luminosity in \citet{rakshit21}. Both X-ray Eddington ratio estimations show a similar positive correlation with the optical Eddington ratios \edd. However, $\lambda_{\rm Edd,X2}$, which uses a luminosity-dependent correction factor, tends to be systematically slightly higher than $\lambda_{\rm Edd,X1}$, where a constant correction factor of $\kappa=20$ is used. The optical Eddington ratios typically fall between the two estimations {(see Table\,\ref{tab_src} for their optical Eddington ratios)}.

In particular, the dashed line in the left panel of Fig. \ref{pic_kappa} represents the best-fit correlation between $\lambda_{\rm Edd,X2}$ and $\lambda_{\rm Edd}$. It shows that when \edd\ is lower than $\log(\lambda_{\rm Edd})=-0.1$, $\lambda_{\rm Edd,X2}$ is lower than $\lambda_{\rm Edd}$, while when \edd\ is higher than this value, $\lambda_{\rm Edd,X2}$ is higher than \edd. The dash-dotted line in the same figure represents the best-fit correlation between \eddxone\ and \edd, from which we find that \eddxone\ {systematically} underestimates the value.

To provide a qualitative assessment, we can calculate the average difference between X-ray and optical Eddington ratios for our sample of NLS1s, assuming the optical Eddington ratios are the `true' values. The average difference is defined as $\langle{\lambda_{\rm Edd,X}-\lambda_{\rm Edd}}\rangle$. We can use the $\chi^{2}$ statistic to quantify the difference as $\Sigma \frac{(\lambda_{\rm Edd,X}-\lambda_{\rm Edd})^2}{\sigma^{2}}$, where $\sigma$ represents the measurement uncertainty of the X-ray Eddington ratios.

The average difference between $\lambda_{\rm Edd,X1}$ and $\lambda_{\rm Edd}$ is -0.03 with $\chi^{2}=24$, and between $\lambda_{\rm Edd,X2}$ and $\lambda_{\rm Edd}$ is 0.03 with $\chi^{2}=13$. Overall, $\lambda_{\rm Edd,X1}$ slightly underestimates the Eddington ratios compared to the optical measurements, while $\lambda_{\rm Edd,X2}$ does the opposite. When using a luminosity-dependent correction factor, $\lambda_{\rm Edd,X2}$ provides a more consistent Eddington ratio with $\lambda_{\rm Edd}$ than $\lambda_{\rm Edd,X1}$ based on the $\chi^{2}$ statistic.

\subsection{X-ray bolometric luminosity correction factors}

Relationships between past and local AGN activity provides important insights into the accretion history of SMBHs. A knowledge of the hard X-ray bolometric luminosity correction factor $\kappa$ is a vital input into these studies, e.g., with the X-ray background \citep{hasinger04,fabian04}. 

Since \eddxtwo\ suggests a slightly higher value than \eddxone, it implies that most of the NLS1s in our sample may have an X-ray bolometric luminosity correction factor $\kappa>20$, which is the typical value for sources with $L_{\rm bol}<10^{46}$ erg s$^{-1}$ \citep{duras20}. We can calculate the X-ray bolometric luminosity using the measured 2-10 keV luminosity and assuming the optical Eddington ratios are accurate. We find that the less luminous X-ray NLS1s with $L_{\rm 2-10 keV}<2\times10^{44}$ erg s$^{-1}$ in our sample require a significantly higher X-ray correction factor than the values calculated in \citet{netzer19}. On the other hand, the more luminous X-ray NLS1s in our sample require a lower X-ray correction factor than ones calculated in \citet{netzer19}. See the purple circles in the right panel of Fig. \ref{pic_kappa} for observational measurements of $\kappa$ and solid black line for the calculations in \citet{netzer19}. This aligns with our previous finding of $\lambda_{\rm Edd,X2}$ showing a steeper correlation with \edd\ than {a} \eddxtwo=\edd\ trend (see the purple dashed and black solid lines in the left panel).

In conclusion, we have demonstrated that by applying a correction factor to X-ray luminosity, we can reproduce similar Eddington ratios as optical measurements. Despite the large measurement uncertainty, primarily due to the limited signal-to-noise in the hard X-ray band of our data, we find tentative evidence suggesting that the X-ray bolometric luminosity correction factor may need revision for our high-\edd\ NLS1s compared to the correction factors  {inferred by} previous work.  In the calculations in \citet{netzer19}, the AGN spectral energy distribution (SED) is described as follows:

\begin{itemize}
    \item The UV and optical emission is dominated by a geometrically thin and optically thick accretion disc model described in \citet{shakura73}, but with full relativistic corrections.
    \item Most recent studies have accounted for variations in bolometric correction using a re-scalable template SED based on the \aox\ or $L_{\rm 2 keV}$-$L_{2500 \AA}$ relation \citep[e.g.,][]{marconi04,hopkins07,netzer19}.
    \item The X-ray portion of the SED is a power law emission with a fixed photon index of $\Gamma=1.9$ in \citet{netzer19}. The normalisation of this X-ray power-law component is determined by the monochromatic luminosity correlation between $L_{\rm 2 keV}$ and $L_{2500 \AA}$ \citep{lusso16}. 
\end{itemize}

It is also important to note that, as \citet{hopkins07} pointed out, the intrinsic scatter in the $L_{\rm 2 keV}$-$L_{2500 \AA}$ relation or the scaling factors give rise to variation of a factor of 2 in the bolometric luminosity correction factor $\kappa$.

We argue that these assumptions used in the estimation of $\kappa$ may not apply to our NLS1s for the following reasons:

\begin{itemize}
    \item Our objects are accreting close to or even above the Eddington limit. The disc may not remain geometrically thin in such extreme accretion regimes \citep[e.g.,][]{skadowski16}.
    \item Our objects exhibit significantly softer X-ray continuum emission than a power law with $\Gamma=1.9$. When fitting the full band data with only a power law, we find $\Gamma\gtrsim2.5$ for most of our objects if we ignore the systematic effects of the soft excess emission as in the SED models of \citet{netzer19}. Therefore, for the same monochromatic luminosity $f_{\rm 2 keV}$, a softer power law leads to a lower 2-10 keV luminosity.
    \item When the soft excess emission is the most significant for the high \edd\ objects, similar to previous \erosita\ study of type-1 AGNs \citep{waddell23}, both soft excess emission and the power law emission contribute to the 2 keV monochromatic luminosity.
    \item Additionally, we cannot rule out the possibility of redshift evolution in $\kappa$, although so far, we have not found significant evidence of different behaviours in our $z\approx1$ NLS1s compared to local NLS1s.
\end{itemize}

Finally, we show the comparison with other theoretical X-ray bolometric luminosity correction factors $\kappa$ as a function of hard X-ray luminosity calculated by \citet{marconi04,hopkins07} as well as observational measurements of AGNs at $z<0.3$ in \citet{vasudevan07} in Fig. \ref{pic_kappa}. The observational measurements of $\kappa$ in \citet{vasudevan07} were done through multi-wavelength SED modelling, a more appropriate {method} than ours which assumes the accuracy of the bolometric luminosity estimation using $L_{3000 \AA}$. Nevertheless, \citet{vasudevan07} also found a significant scatter in the $\kappa$-$L_{\rm 2-10 keV}$ correlation and emphasised the importance of simultaneous multi-wavelength data in measuring $\kappa$ because of the known intrinsic UV and X-ray variability, which we do not have as the SDSS and \erosita\ data are not simultaneous.

\begin{table}
    \centering
        \caption{The 2-10 keV unabsorbed luminosity and X-ray Eddington ratios. $\lambda_{\rm Edd, X1}$ is based on an X-ray bolometric luminosity correction factor of 20; $\lambda_{\rm Edd, X2}$ is based on an X-ray luminosity-dependent correction factor \citep{netzer19}; $\kappa$ is the X-ray correction factor assuming the optical Eddington ratio $\lambda_{\rm Edd}$ in Table \ref{tab_src} is accurate and precise.}
    \label{tab_eddington}
    \begin{tabular}{ccccc}
    \hline\hline
       Names  & $L_{\rm X}$ (2-10 keV) & $\log(\lambda_{\rm Edd, X1})$ & 
    $\log(\lambda_{\rm Edd, X2})$ & $\kappa$\\
    & $10^{44}$ erg s$^{-1}$ \\
       \hline
J0235 & $5.3\pm4.1$ & $-0.4\pm0.4$ &  $0.0\pm0.5$ & $45$ \\
J0824 & $11.9\pm6.4$ & $0.1\pm0.2$ &  $0.5\pm0.3$ & $25$ \\
J0826 & $6.9\pm4.5$ & $0.2\pm0.4$ &  $0.6\pm0.4$ & $22$ \\
J0845 & $1.7\pm1.5$ & $-0.8\pm0.3$ &  $-0.6\pm0.5$ & $116$ \\
J0859 & $4.5\pm2.5$ & $-0.1\pm0.5$ &  $0.2\pm0.5$ & $19$ \\
J0921 & $1.0\pm1.7$ & $-0.9\pm0.5$ &  $-0.7\pm0.7$ & $94$ \\
J0940 & $10.0\pm5.5$ & $-0.2\pm0.3$ &  $0.3\pm0.3$ & $36$ \\
J0957 & $4.3\pm2.5$ & $-0.2\pm0.4$ &  $0.2\pm0.4$ & $23$ \\
J1026 & $1.2\pm2.1$ & $-0.8\pm0.6$ &  $-0.6\pm0.7$ & $58$ \\
J1036 & $6.0\pm4.4$ & $0.1\pm0.4$ &  $0.5\pm0.4$ & $24$ \\
J1045 & $10.0\pm7.6$ & $0.1\pm0.4$ &  $0.5\pm0.4$ & $16$ \\
J1133 & $3.2\pm2.3$ & $-0.6\pm0.3$ &  $-0.3\pm0.4$ & $89$ \\
J1148 & $2.4\pm2.8$ & $-0.2\pm0.5$ &  $0.0\pm0.5$ & $11$ \\
J1304 & $1.3\pm0.9$ & $-0.7\pm0.4$ &  $-0.5\pm0.4$ & $67$ \\
\hline\hline
    \end{tabular}
\end{table}

\section{Future Work}

\subsection{The need for a deeper X-ray survey and an appropriate statistical approach}

Firstly, our analysis of the flux-limited sample highlights a trend: NLS1s at $z\approx1$ exhibit a similar \aox\ as local AGNs with similar UV luminosities. We specifically selected the {sources} with the highest X-ray detection likelihood. 

While our current dataset does not provide sufficient statistical power to establish a correlation between UV and X-ray monochromatic luminosities, the prospect of deeper X-ray surveys holds promise for elucidating such relationships. Understanding the redshift evolution of \aox\ is vital, as it sheds light on variations in accretion physics, including the formation of X-ray coronae, across different cosmic epochs. Given the known correlations between $L_{2500 \AA}$, $L_{\rm 2 keV}$, and \aox, disentangling the true underlying drivers presents a challenge. Moreover, it remains unclear whether the redshift evolutions of $L_{\rm 2 keV}$ and \aox\ are inherently linked, as discussed in Section 3.6 of \citet{steffen06}.

To address these complexities, future endeavours should not only entail deeper surveys of high-redshift AGNs and quasars but also employ robust statistical methodologies. The Random Forest approach, for instance, offers a systematic means of exploring intricate parameter correlations, facilitating a deeper understanding of the relationship within multi-parameter space \citep[e.g., see an application in][]{piotrowska21}.

\subsection{Search for high-$z$, X-ray-weak AGNs} \label{sec_xrayweaknls1}

\subsubsection{The current and future X-ray surveys for $z\gtrsim1$ NLS1s}

All of our $z\approx1$ NLS1s exhibit consistency with the expected 2 keV monochromatic luminosity given their 2500 \AA\ monochromatic luminosity, as per the correlation derived from AGNs spanning $z=0.04-4.25$ \citep{lusso10}. Alongside their soft X-ray continuum emission measured by \erosita, they likely resemble the `simple' X-ray NLS1s observed in the local Universe. 

However, in previous work, we identified the NLS1 J0334 at $z=0.35$ exhibiting a weak X-ray state, notably falling well below the $L_{\rm 2 keV}$ vs $L_{2500 \AA}$ correlation \citep{yu23}. This prompts the question of how many `complex' X-ray NLS1s we might uncover at $z\gtrsim1$, similar to the ones in the local X-ray Universe that manifest X-ray weak states \citep[e.g.,][]{parker14}. Addressing this query would necessitate a survey with a signal-to-noise ratio comparable to eRASS1 {but with an order of magnitude improvement in flux sensitivity}.

In the right panel of Fig. \ref{pic_dis}, we show constant flux levels of $10^{-12}$, $10^{-14}$ and $10^{-16}$ \ergps, with each point along a given curve corresponding to the same observed flux\footnote{{It is} essential to note that the constant-flux curves in Fig. \ref{pic_dis} serve as benchmarks for the detector's sensitivity, demonstrating the luminosity levels detectable by an X-ray survey at different redshifts or cosmic epochs. However, {it is} crucial to acknowledge that the detector's sensitivity is not uniform across varying redshifts. This discrepancy arises because the rest-frame hard X-ray emission undergoes redshifting, shifting into softer X-ray bands of the detector, thereby leading to changes in sensitivity levels.}. Future medium and deep X-ray surveys, exemplified by telescopes like \textit{AXIS} \citep{reynolds23} and \textit{NewAthena} \citep{cruise25}, hold promise for exploring even fainter objects. {These upcoming surveys have the potential to uncover lower-mass BHs or lower-X-ray luminosity NLS1 AGNs beyond the local X-ray Universe and provide new insights into more extreme accretion regimes.} 

{For example, a 15~Ms \textit{AXIS} survey is projected to achieve a flux sensitivity of $10^{-17}$ \ergps\ in a wide-field survey configuration (a single tile exposure of 15\,ks covering 50 deg$^{2}$) in the observed 0.5–2keV band \citep{marchesi20}. For comparison, our eRASS1 and SDSS sample of $z\approx1$ NLS1s has an average observed flux of $10^{-14}-10^{-13}$ \ergps\ in the observed 0.5--2\,keV band. This suggests that an \textit{AXIS} survey in such a configuration will be capable of detecting SDSS-observed NLS1s at $z\approx1$ with comparable rest-frame 2500~\AA\ luminosity but X-ray luminosities at least an order of magnitude lower or more—corresponding to \aox\ $<-2.5$. Such sensitivity would enable the detection of the most extreme X-ray-weak NLS1s, such as PHL 1092, in its lowest observed X-ray luminosity state \citep{miniutti09}, but at much greater cosmic distances.}

\subsubsection{The number density of X-ray-weak AGNs}

{With the advent of deeper X-ray surveys of X-ray-weak AGNs, studying the evolution of \aox\ as a function of redshift may become feasible. Fig.\,\ref{pic_xuv} compares the UV and X-ray monochromatic luminosities of our $z\approx1$ NLS1s with those of $z>4$ AGNs from \citet{steffen06} and $z>4$ quasars from \citet{steffen06,nanni17}. Notably, radio-loud quasars—typically more luminous than their radio-quiet counterparts across wavelengths—exhibit smaller \aox\ values at higher redshifts, implying relatively weaker X-ray luminosity \citep[e.g.,][]{yue24,kocevski24}. While some studies suggest that \aox\ evolves with redshift and exhibits a luminosity dependence, others find no significant trend \citep[see Sections 3.5--3.6 in][for a more detailed discussion]{steffen06}.}

{The cosmic evolution of X-ray-weak AGNs remains an open question, particularly in light of the high number densities of these broad-line AGNs observed by JWST (see Section\,\ref{sec_uvx}). These AGNs appear to contribute significantly to the cosmic photoionisation rate-potentially up to 50 per cent at $z\approx6$-comparable to X-ray-selected AGNs \citep{harikane23}. While such optical-red, UV-blue and X-ray-weak AGNs are commonly detected at $z\approx5-6$, they may also exist at even higher redshifts, as hinted by some earlier studies \citep[e.g.,][]{fujimoto22,endsley23}.}

{We refrained from using our flux-limited sample of NLS1s at $z\approx1$ to establish any statistical correlation between UV and X-ray monochromatic luminosity. However, it is conceivable that such correlations could be explored in the future with deeper X-ray surveys.}

\subsubsection{The duty cycle of the weak X-ray state}

{Observational studies of local NLS1s indicate that these AGNs, particularly those classified as ‘complex’ NLS1s in \citet{gallo06}, exhibit extreme X-ray variability. As discussed in Section \ref{sec_uvx}, the \aox\ values of these NLS1s can deviate significantly from the expected average based on their UV luminosity, shifting to much lower X-ray luminosities over timescales of months \citep[e.g.,][]{tripathi20} to years \citep[e.g.,][]{grupe00}. Detailed X-ray studies attribute this variability primarily to changes in the intrinsic X-ray emission, implying changes in the innermost accretion geometry \citep{miniutti09,parker14,jiang18}. It is also worth noting that variable absorption features have been observed in combination with the changes in primary X-ray emission \citep[e.g., {in the `complex' NLS1s 1H\,0707$-$495 and Mrk\,335,}][]{boller21, liu21}. The duty cycle of such weak X-ray luminosity state varies across different NLS1s. For example, Mrk~335 remained in a weak X-ray luminosity state (\aox\ < -1.8) for approximately 3 out of 13 years between 2007 and 2020 \citep[e.g.,][]{tripathi20,kara23}. } 

{In addition to intrinsic X-ray variability, NLS1s also exhibit extreme transient events, where multi-wavelength variability indicates significant changes in their accretion rates. For example, the NLS1 AT2021aeuk, as studied in \citet{sun25}, underwent a dramatic change in luminosity. A multi-wavelength monitoring program revealed that its X-ray luminosity decreased by a factor of 100 within 3 months, while its optical luminosity peaked. }

{Similar X-ray-weak states have been observed in other types of AGNs undergoing transient events, as mentioned in Section \ref{sec_uvx} \citep[e.g.,][]{ricci21, payne23}. For instance, the 0.3–10 keV X-ray luminosity of the bare Seyfert 2 AGN 1ES~1927$+$654  dropped to a low value of $10^{42}$ erg s$^{-1}$ soon after its optical luminosity peaked \citep{masterson22}. Within the following year, the X-ray luminosity increased again to approximately $10^{44}$ erg s$^{-1}$, reaching the Eddington luminosity for a $10^{6}M_{\odot}$ black hole. The dramatic drop and subsequent reappearance of X-ray luminosity in both AT2021aeuk and 1ES~1927$+$654 are often interpreted as tidal disruption events, where the X-ray-emitting coronal region is destroyed and later restored \citep{ricci21,payne23}.}

{Alternatively, the high \aox\ value of 1ES1927$+$654 (\aox$=-1$) during its pre-transient state in 2011 \citep{laha22} motivated the models of magnetically arrested disks in \citet{scepi21}, where the disappearance of X-ray emission during the transient event is explained as a reset of the magnetic field that powers the coronal region.}

{How often does such a significant drop in X-ray luminosity occur? The sample remains small. Among the objects we already know are most likely repeating tidal disruption events, the NLS1 AT2021aeuk has experienced two episodes separated by 2.9 years \citep{sun25}, and ASASSN-14ko has undergone six episodes with a period of around 0.31 years \citep{payne23}. These peculiar transient events raise questions about the duty cycle of the weak X-ray states, which is closely related to the detection possibility of X-ray-weak NLS1s or generally AGNs in the distant Universe.}

\subsection{X-ray bolometric luminosity correction factor for high-Eddington ratio NLS1s}

Finally, while acknowledging the substantial measurement uncertainty primarily from the limited signal-to-noise in the hard X-ray band of our data, we tentatively observe that the X-ray bolometric luminosity correction factor may require adjustment for our samples compared to the factors calculated in \citet{netzer19}. We posit that certain assumptions underlying the geometrically thin accretion disc models and the only power-law model used for the X-ray portion of the SED template in previous theoretical calculations of X-ray bolometric luminosity correction may not be applicable to our near or even super-Eddington NLS1 AGNs.

With the discovery of more AGNs in the early Universe \citep[$z\gtrsim10$,][]{kovacs24}, an X-ray bolometric luminosity correction factor derived from local correlations has been employed to estimate bolometric luminosity and thus BH masses assuming {Eddington-limited accretion}. We {urge} caution using such a luminosity correction factor without due consideration of the unique characteristics of these AGNs, {the need for which is}  evidenced by our $z\approx1$ samples of a similar luminosity.  {The application of such a standard X-ray bolometric luminosity correction factor requires particular caution for `complex' NLS1s, as they frequently exhibit X-ray weak states with lower \aox\ values compared to ‘simple’ NLS1s and BLS1s \citep{gallo06}. In addition, as discussed in Section~\ref{sec_xrayweaknls1}, `complex' NLS1s often display a strong soft excess, which contributes significantly to the soft X-ray emission but has yet to be accounted in the calculations of these standard X-ray bolometric luminosity correction factors.}

\section{Conclusions}

We conducted a cross-search between the SDSS-observed NLS1s and the eRASS1 catalogue, identifying 14 X-ray sources associated with NLS1s at $z\approx1$. These NLS1s represent a unique stage in the BH growth history, characterised by low $m_{\rm BH}$ and high \edd\ values.

First, all of the $z\approx1$ NLS1s exhibit agreement with  the expected 2 keV monochromatic luminosity given their 2500 \AA\ monochromatic luminosity, as derived from correlations established in previous studies on AGNs spanning $z=0.04$ to $4.25$ \citep{lusso10}. This suggests a similarity to the `simple' X-ray NLS1s observed in the local Universe, as previously classified in \citet{gallo06}, contrasting with `complex' X-ray NLS1s that {typically} exhibit X-ray weak states.

Second, the majority of the very soft X-ray continuum emissions, when fitted with a power-law model, necessitate photon indices $\Gamma\gtrsim2.5$, with the exception of three sources consistent with $\Gamma=2$. {The median photon index of our sample of $z\approx1$ NLS1s is 2.7, which is consistent with the median value found in a much larger sample of \textit{eROSITA}-observed NLS1s at $z\lessapprox0.8$ \citep{grunwalk23}.} Notably, the highest photon index, observed in J0845, hints at a significant contribution from soft excess emission within the energy band covered by \erosita. 

Finally, our analysis demonstrates that we can align the Eddington ratios with optical measurements by applying a correction factor to the X-ray luminosity. Although measurement uncertainty remains considerable, primarily due to limited signal-to-noise in the hard X-ray band, we suggest that the X-ray bolometric luminosity correction factor may need adjustments for our high-\edd\ objects compared to theoretical values calculated from rescaled AGN templates assuming the standard thin disc model. We argue that certain assumptions underlying previous estimations of X-ray bolometric luminosity correction, such as those related to geometrically thin accretion disc models and hard power law models for the X-ray continuum, may not hold for our near or even super-Eddington NLS1 AGNs. 

\section*{Acknowledgements}

J.J. gratefully acknowledges Jeremy Sanders for his invaluable assistance in understanding \erosita\ data and guiding the data reduction processes. {We thank the anonymous reviewer for their careful reading of our manuscript and their many insightful comments and suggestions.} {DJW acknowledges support from the Science and Technology Facilities Council
(STFC; grant code ST/Y001060/1).} This work is based on data from \erosita, the soft X-ray instrument aboard SRG, a joint Russian-German science mission supported by the Russian Space Agency (Roskosmos), in the interests of the Russian Academy of Sciences represented by its Space Research Institute (IKI), and the Deutsches Zentrum für Luft- und Raumfahrt (DLR). The SRG spacecraft was built by Lavochkin Association (NPOL) and its subcontractors and is operated by NPOL with support from the Max Planck Institute for Extraterrestrial Physics (MPE). The development and construction of the eROSITA X-ray instrument were led by MPE, with contributions from the Dr. Karl Remeis Observatory Bamberg \& ECAP (FAU Erlangen-Nuernberg), the University of Hamburg Observatory, the Leibniz Institute for Astrophysics Potsdam (AIP), and the Institute for Astronomy and Astrophysics of the University of Tübingen, with the support of DLR and the Max Planck Society. The Argelander Institute for Astronomy of the University of Bonn and the Ludwig Maximilians Universität Munich also participated in the science preparation for \erosita.

\section*{Data Availability}

All the data can be downloaded from the eRASS1 website at https://erosita.mpe.mpg.de/dr1/. A machine-readable file in FITS format containing Tables \ref{tab_src}, \ref{tab_flux}, \ref{tab_pl}, \ref{tab_eddington} and \ref{tab_mag} is available for download at https://github.com/jcjiang-dev/erassnls1. 
 



\bibliographystyle{mnras}
\bibliography{example} 




\appendix

\section{Optical and Ultraviolet magnitudes of the NLS1s in our sample}

\begin{table}
    \centering
    \begin{tabular}{cccc}
    \hline\hline
    Names & NUV & $i$ & W2\\
    \hline
      J0235  &  $20.04\pm0.03$ & $19.32\pm0.04$ & $14.59\pm0.14$ \\
      J0235  &  $-$ & $17.651\pm0.007$ & $12.71\pm0.03$ \\
      J0826  &  $20.13\pm0.10$ & $18.54\pm0.02$ & $14.23\pm0.05$ \\
      J0845  &  $-$ & $18.23\pm0.04$ & $-$ \\
      J0859  &  $19.26\pm0.09$ & $19.058\pm0.018$ & $14.10\pm0.11$ \\
      J0921  &  $19.73\pm0.09$ & $18.802\pm0.017$ & $13.95\pm0.04$ \\
      J0940  &  $18.73\pm0.05$ & $17.777\pm0.009$ & $13.22\pm0.03$ \\
      J0957  &  $19.30\pm0.05$ & $19.05\pm0.02$ & $13.90\pm0.05$ \\
      J1026  &  $20.28\pm0.16$ & $19.83\pm0.03$ & $14.19\pm0.04$ \\
      J1036  &  $19.28\pm0.09$ & $18.624\pm0.018$ & $13.43\pm0.07$ \\
      J1045  &  $20.34\pm0.06$ & $19.20\pm0.03$ & $14.51\pm0.06$ \\
      J1133  &  $19.33\pm0.03$ & $18.243\pm0.015$ & $13.50\pm0.03$ \\
      J1148  &  $21.5\pm0.3$ & $19.91\pm0.03$ & $14.02\pm0.09$ \\
      J1304  &  $19.28\pm0.05$ & $19.03\pm0.03$ & $14.62\pm0.10$ \\
      \hline\hline
    \end{tabular}
    \caption{Magnitudes of our samples in NUV measured by \textit{GALEX}, $i$ measured by SDSS and W2 measured by \textit{WISE}. The values are from NED \citep{cook23}.}
    \label{tab_mag}
\end{table}

AB magnitudes of our samples in the UV, optical and near-infrared wavelengths are shown in Table\,\ref{tab_mag} for readers' reference. Interested readers may refer to NED \citep{cook23} for magnitudes {at} other wavelengths. 

\section{SDSS and eRASS-inferred coordinate difference} \label{sec_cor}

\begin{figure}
    \centering
    \includegraphics[width=7cm]{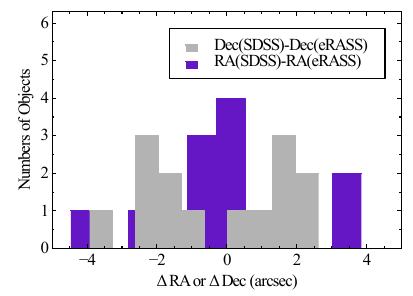}
    \caption{The SDSS and eRASS RA and Dec coordinate difference of our targets.}
    \label{pic_ra_dec}
\end{figure}

{Fig.\,\ref{pic_ra_dec} shows the difference between the RA and Dec coordinates of the NLS1 AGNs inferred by SDSS and eRASS. J1148 and J1045 are the only two objects with a difference in RA coordinates greater than 3.6 arcsec, while J1304 is the only object with a DEC coordinate difference exceeding 3.6 arcsec. The other targets all have a coordinate difference smaller than 3.6 arcsec.}

\section{Source and background regions for spectral extraction}

This section includes supplementary tables and figures for the data analysis. Table\,\ref{tab_regions} provides the sizes of the source and background regions used for spectral extraction. Figures\,\ref{pic_ap1}--\ref{pic_ap3} display the spectra of our targets alongside their best-fit models.

\begin{table}
    \centering
    \begin{tabular}{cccc}
    \hline\hline
     Names   & Source Regions & \multicolumn{2}{c}{Background Regions} \\
                  & $\times10^{-2}$   & $\times10^{-2}$ & $\times10^{-1}$ \\
                  & degrees & degrees & degrees \\
                  \hline
J0235&1.39&3.17&1.72\\
J0824&1.92&4.17&2.38\\
J0826&1.64&3.61&2.04\\
J0845&1.92&4.11&2.38\\
J0859&1.72&3.89&2.14\\
J0921&1.56&3.44&1.94\\
J0940&1.83&4.00&2.28\\
J0957&1.64&3.78&2.04\\
J1026&1.56&3.44&1.94\\
J1036&1.56&3.44&1.94\\
J1045&1.53&3.44&1.90\\
J1133&1.56&3.44&1.94\\
J1148&1.44&3.33&1.80\\
J1304&1.44&3.28&1.80\\
\hline\hline
    \end{tabular}
    \caption{The sizes of the source and background regions to extract data products. The second column is the radii of the circular source regions. The third and fourth columns are the inner and outer radii of the annulus background regions.}
    \label{tab_regions}
\end{table}

\begin{figure*}
    \centering
    \includegraphics[width=\textwidth]{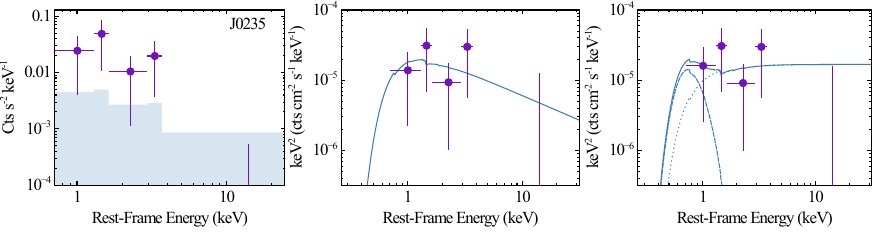}
    \includegraphics[width=\textwidth]{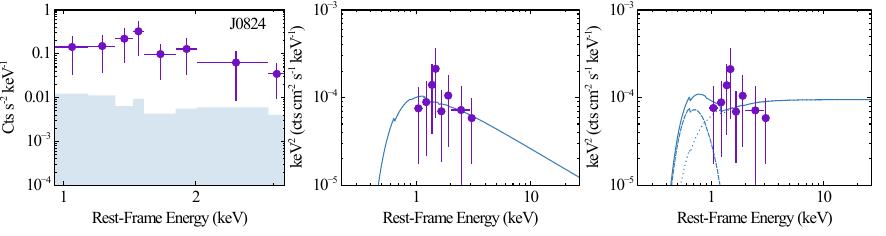}
    \includegraphics[width=\textwidth]{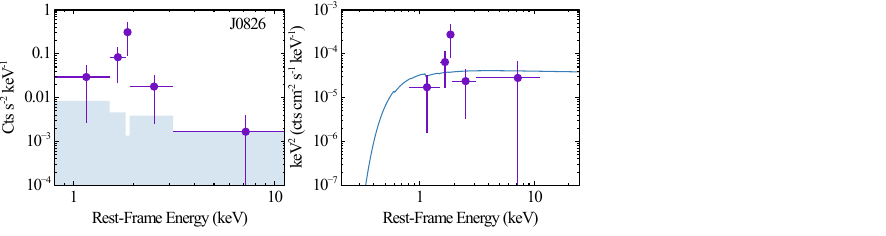}
    \includegraphics[width=\textwidth]{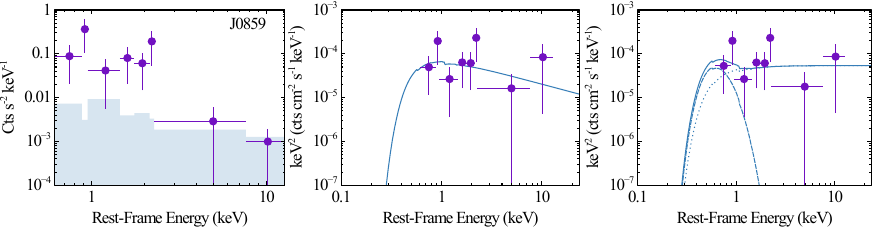}
    \caption{Same as Fig. \,\ref{pic_model} but for other NLS1s in the sample.}
    \label{pic_ap1}
\end{figure*}

\begin{figure*}
    \centering
    \includegraphics[width=\textwidth]{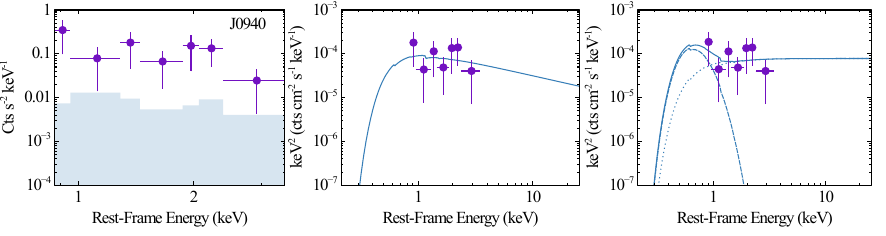}
    \includegraphics[width=\textwidth]{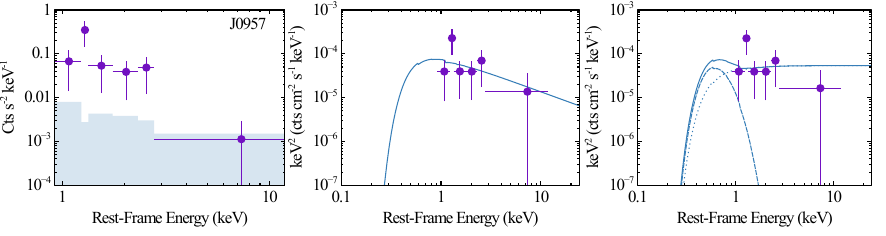}
    \includegraphics[width=\textwidth]{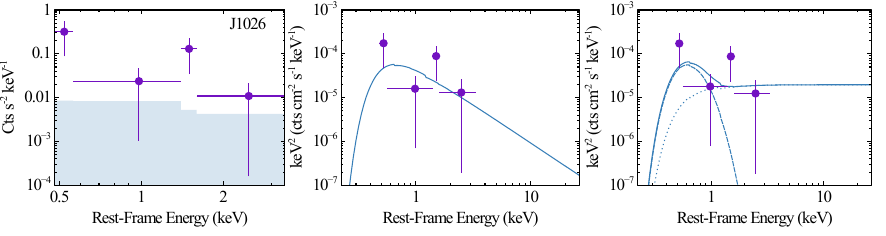}
    \includegraphics[width=\textwidth]{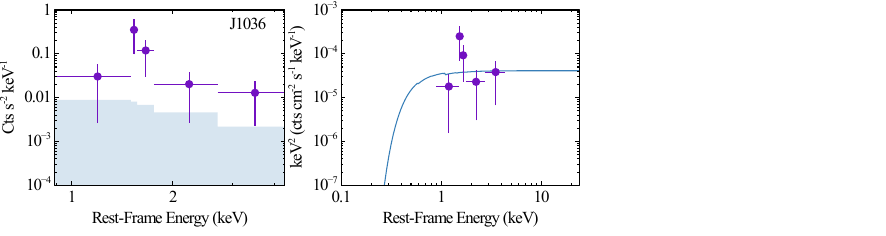}
    \caption{Continued.}
    \label{pic_ap2}
\end{figure*}

\begin{figure*}
    \centering
    \includegraphics[width=\textwidth]{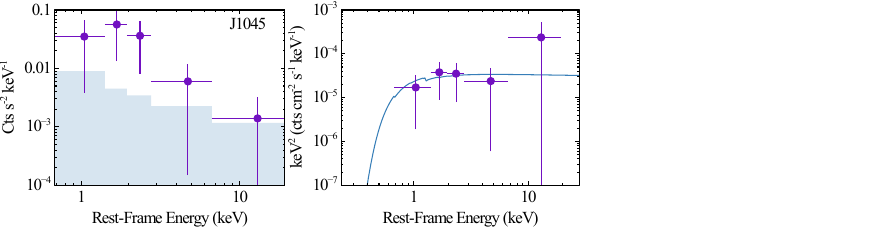}
    \includegraphics[width=\textwidth]{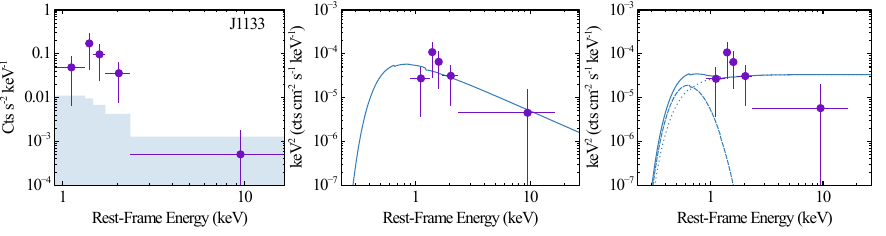}
    \includegraphics[width=\textwidth]{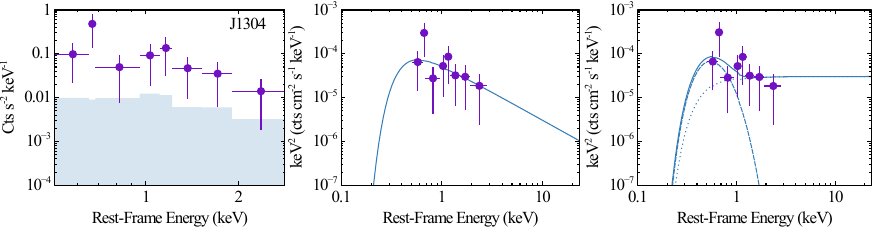}
    \caption{Continued.}
    \label{pic_ap3}
\end{figure*}

\section{Modelling the background spectra} \label{sec_bkg_model}

In this section, we fit the \erosita\ background spectrum obtained from the background {region, rather than simply subtracting it from the source spectrum, and include this as an additional component when modelling the spectra from the source region}. This approach is necessary due to the limited number of net counts for our targets in eRASS1. 

\begin{figure}
    \centering
    \includegraphics[width=7cm]{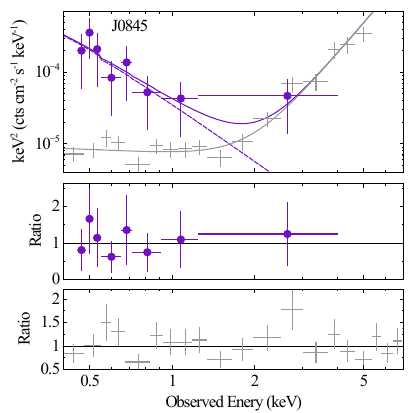}
    \caption{Top: Unfolded total spectrum (purple cross bars) and background spectrum (grey cross bars) of J0845, along with their corresponding best-fit models shown in purple and grey, respectively. The purple dashed line represents the source contribution, modelled by an absorbed power law. Middle: Data-to-model ratio for the total spectrum. Bottom: Data-to-model ratio for the background spectrum.}
    \label{pic_back}
\end{figure}

We use J0845 as a demonstration case. The background spectrum was modeled using two power-law components, and the best-fit model, along with the unfolded background spectrum, is shown in Fig.\,\ref{pic_back}. The best-fit power law photon indices for the background spectrum are 2.2 and -2.6. The same model was then applied when fitting the source spectrum. The purple dashed line represents the absorbed power-law component for the source, while the purple solid line shows the total model, which includes both the best-fit source and background components. The resulting source power law has a photon index of $\Gamma=4.7\pm1.1$, with C-stat/$\nu$ = 1.79/6, consistent with the results presented in Table\,\ref{tab_pl}. We {found} similar conclusions for other objects. Background and source model parameters are presented in Table\,\ref{tab_back}.

\begin{table}
    \centering
    \begin{tabular}{c|ccccc}
    \hline\hline
      Names & $\Gamma_{\rm bkg,1}$ & $\Gamma_{\rm bkg,2}$ & C-stat/$\nu$ & $\Gamma_{\rm src}$ & C-stat/$\nu$  \\
      \hline
       J0235  &  $-2.7\pm0.3$ & $1.7\pm0.3$ & 84.21/62 & $2.6\pm1.0$ & $2.02/3$ \\
       J0824  &  $-2.4\pm0.4$ & $1.9^{+0.5}_{-0.4}$ & 70.88/63 & $2.8\pm0.4$ & $4.93/6$ \\
       J0826  &  $-2.5\pm0.4$ & $1.9\pm0.3$ & 44.12/57 & $2.0\pm0.5$ & $6.10/3$ \\
       J0845  &  $-2.6^{+0.4}_{-0.3}$ & $2.2\pm0.2$ & 75.41/74 & $4.7\pm1.1$ & $1.79/6$ \\
       J0859  &  $-2.4\pm0.3$ & $2.3\pm0.5$ & 100.84/83 & $2.8\pm0.6$ & $10.34/6$ \\
       J0921  &  $-2.4\pm0.5$ & $2.0^{+0.5}_{-0.4}$ & 42.43/43 & $3.0\pm0.4$ & $0.30/2$ \\
       J0940  &  $-2.4\pm0.3$ & $2.0^{+0.4}_{-0.3}$ & 64.15/66 & $2.6\pm0.7$ & $4.76/5$ \\
       J0957  &  $-2.2^{+0.3}_{-0.4}$ & $2.2^{+0.6}_{-0.5}$ & 73.81/59 & $3.2\pm0.4$ & $5.02/4$ \\
       J1026  &  $-2.6^{+0.4}_{-0.3}$ & $2.0\pm0.3$ & 33.71/54 & $3.7\pm1.1$ & $6.42/2$ \\
       J1036  &  $-2.8\pm0.3$ & $1.9^{+0.3}_{-0.4}$ & 69.36/58 & $1.8\pm1.0$ & $6.02/3$ \\
       J1045  &  $-2.2\pm0.4$ & $2.4^{+0.6}_{-0.5}$ & 70.28/47 & $1.8\pm1.0$ & $1.02/3$ \\
       J1133  &  $-2.6\pm0.4$ & $2.3\pm0.3$ & 68.27/56 & $3.1\pm1.1$ & $3.01/3$ \\
       J1148  &  $-2.0\pm0.3$ & $2.8\pm0.5$ & 72.11/47 & $2.3^{+0.7}_{-0.8}$ & $1.74/3$ \\
       J1304  &  $-2.7\pm0.4$ & $2.3\pm0.2$ & 97.56/81 & $3.4\pm0.7$ & $4.42/6$ \\
    \hline\hline
    \end{tabular}
    \caption{The best-fit power-law photon index ($\Gamma_{\rm bkg,1}$, $\Gamma_{\rm bkg,2}$) for the background spectra and corresponding goodness of fit. $\Gamma_{\rm src}$ is the best-fit photon index of the source after fitting the spectra with fixed background models.}
    \label{tab_back}
\end{table}

\section{Soft excess strength} \label{sec_soft}

\begin{figure}
    \centering
    \includegraphics[width=8cm]{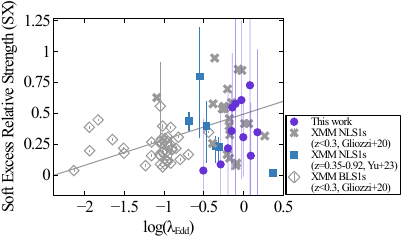}
    \caption{Soft excess strength vs Eddington ratio of NLS1s (purple circles, blue squares and grey crosses) and BLS1s (grey diamonds). The symbols represent the same sample as in Fig. \ref{pic_gamma}. The solid line is the linear regression of the two parameters of the NLS1 and BLS1 samples in \citet{gliozzi20}.}
    \label{pic_sx}
\end{figure}

The very soft continuum observed in our NLS1 sample suggests that soft excess emission may significantly contribute to the total X-ray luminosity. In Section \ref{bbfit}, we attempted to model the soft excess using a simple blackbody component. However, due to the limited signal-to-noise ratio, the spectral shape of the soft excess could not be well constrained. We therefore assumed a blackbody temperature of 0.1 keV to attempt to estimate the strength of the soft excess for the objects in the same manner as in \citet{gliozzi20,yu23} for a sensible comparison. We define the soft excess strength $SX$ as $SX=\frac{F_{\rm bb}}{F_{\rm pl}}$, where $F_{\rm bb}$ and $F_{\rm pl}$ represent the unabsorbed flux of the blackbody and power-law components, respectively, in the rest frame 0.5-10 keV band. Although we acknowledge the associated uncertainty, the soft excess strength $SX$ of these $z\approx1$ NLS1s is comparable to those observed in the local Universe, as illustrated in Fig. \ref{pic_sx}.

\begin{table}
    \centering
        \caption{Best-fit power law plus blackbody model for our {NLS1s}. The photon index of the power law is fixed at 2. The temperature of the blackbody is fixed 0.1 keV.}
    \label{tab_bb}
    \begin{tabular}{ccccc}
    \hline\hline
    Names & $\log(F_{\rm bb})$ & $\log(F_{\rm pl})$ & SX & C-stat/$\nu$ \\
         & \ergps & \ergps \\
         \hline
J0235 & $-12.69\pm0.25$ & $-13.30\pm0.59$ & $0.6\pm0.6$ & 2.38/3 \\
J0824 & $-12.34\pm0.15$ & $-12.69\pm0.65$ & $0.3\pm0.7$ & 5.55/6 \\
J0845 & $-12.77\pm0.27$ & $-13.08\pm0.17$ & $0.3\pm0.3$ & 2.45/6 \\
J0859 & $-12.59\pm0.17$ & $-13.14\pm0.41$ & $0.6\pm0.4$ & 9.32/6 \\
J0921 & $-12.83\pm0.35$ & $-13.05\pm0.31$ & $0.2\pm0.5$ & 0.09/2 \\
J0940 & $-12.43\pm0.16$ & $-12.59\pm0.40$ & $0.2\pm0.4$ & 4.04/5 \\
J0957 & $-12.59\pm0.18$ & $-13.17\pm0.85$ & $0.6\pm0.9$ & 6.16/4 \\
J1026 & $-13.02\pm0.32$ & $-13.11\pm0.28$ & $0.1\pm0.4$ & 6.26/2 \\
J1133 & $-12.79\pm0.22$ & $-13.52\pm1.08$ & $0.7\pm1.1$ & 4.73/3 \\
J1148 & $-12.84\pm0.26$ & $-12.88\pm0.33$ & $0.0\pm0.4$ & 0.70/3 \\
J1304 & $-12.84\pm0.21$ & $-13.20\pm0.25$ & $0.4\pm0.3$ & 5.21/6 \\
\hline\hline
    \end{tabular}
\end{table}


\bsp	
\label{lastpage}
\end{document}